\documentclass[10pt,journal,compsoc]{IEEEtran}
\ifCLASSOPTIONcompsoc
  \usepackage[nocompress]{cite}
\else
  \usepackage{cite}
\fi
\ifCLASSINFOpdf
\else
\fi

\usepackage{booktabs} % For formal tables
\usepackage[mathscr]{euscript}
\usepackage{wrapfig}
\usepackage{amsfonts}
\usepackage{amssymb}
\usepackage{amsmath}
\usepackage{mathptmx}
\usepackage{amsthm}
\usepackage{graphicx}
\usepackage{color, colortbl}
\definecolor{lightgray}{gray}{0.9}
\usepackage{xcolor}
\usepackage{hyperref}
\hypersetup{
    colorlinks=true,
    citecolor=cyan,
    linkcolor=magenta,
    filecolor=magenta,
    urlcolor=cyan,
}
\hypersetup{draft}

\usepackage[boxed,ruled,linesnumbered]{algorithm2e} % For algorithms

\SetCommentSty{mycommfont}
\let\oldnl\nl% Store \nl in \oldnl
\newcommand{\nonl}{\renewcommand{\nl}{\let\nl\oldnl}}% Remove line number for one line

\SetAlFnt{\small}
\SetAlCapFnt{\small}
\SetAlCapNameFnt{\small}
\SetAlCapHSkip{0pt}
\IncMargin{-\parindent}

\usepackage[english]{babel}

\newcommand{\RNum}[1]{\uppercase\expandafter{\romannumeral #1\relax}}
\newcommand{\figref}[1]{Fig.~\ref{#1}}
\newcommand{\tabref}[1]{Table~\ref{#1}}
\newcommand{\eqnref}[1]{Eq.~\eqref{#1}}

\newcommand{\comments}[1]{}
\newlength\savedwidth
\newcommand\whline[1]{\noalign{\global\savedwidth\arrayrulewidth
                               \global\arrayrulewidth #1} %
                      \hline
                      \noalign{\global\arrayrulewidth\savedwidth}}

\begin{document}
\title{DeepWarp: DNN-based Nonlinear Deformation}

\author{Ran~Luo,~\IEEEmembership{Student~Member,~IEEE,}
        Tianjia~Shao,~\IEEEmembership{Member,~IEEE,}
        Huamin~Wang,~\IEEEmembership{Member,~IEEE,}
        Weiwei~Xu,~\IEEEmembership{Member,~IEEE,}
        Kun Zhou,~\IEEEmembership{Fellow,~IEEE,}
        and~Yin~Yang,~\IEEEmembership{Member,~IEEE}% <-this % stops a space
\IEEEcompsocitemizethanks{\IEEEcompsocthanksitem Ran Luo and Yin Yang are with Department of Electrical and Computer Engineering, University of New Mexico, NM, 87131\protect\\
E-mail: \{luoran$|$yangy\}@unm.edu
\IEEEcompsocthanksitem Tianjia Shao is with the School of Computing, University of Leeds, UK, LS29JT.\protect\\
E-mail:tianjiashao@gmail.com
\IEEEcompsocthanksitem Huamin Wang is with Department of Computer Science and Engineering, Ohio State University, OH, 43210.\protect\\
E-mail: whmin@cse.ohio-state.edu
\IEEEcompsocthanksitem Weiwei Xu and Kun Zhou are with State Key Lab of CAD\&CG at Zhejiang University, China.\protect\\
E-mail: weiwei.xu.g@gmail.com;kunzhou@acm.org

}
}% <-this % stops an unwanted space
\IEEEtitleabstractindextext{%
\begin{abstract}
DeepWarp is an efficient and highly re-usable deep neural network (DNN) based nonlinear deformable simulation framework. Unlike other deep learning applications such as image recognition, where different inputs have a uniform and consistent format (e.g. an array of all the pixels in an image), the input for deformable simulation is quite variable, high-dimensional, and parametrization-unfriendly. Consequently, even though DNN is known for its rich expressivity of nonlinear functions, directly using DNN to reconstruct the force-displacement relation for general deformable simulation is nearly impossible. DeepWarp obviates this difficulty by partially restoring the force-displacement relation via warping the nodal displacement simulated using a simplistic constitutive model -- the linear elasticity. In other words, DeepWarp yields an incremental displacement fix based on a simplified (therefore incorrect) simulation result other than returning the unknown displacement directly. We contrive a compact yet effective feature vector including \emph{geodesic}, \emph{potential} and \emph{digression} to sort training pairs of per-node linear and nonlinear displacement. DeepWarp is robust under different model shapes and tessellations. With the assistance of deformation substructuring, one DNN training is able to handle a wide range of 3D models of various geometries including most examples shown in the paper. Thanks to the linear elasticity and its constant system matrix, the underlying simulator only needs to perform one pre-factorized matrix solve at each time step, and DeepWarp is able to simulate large models in real time.\end{abstract}

}

\maketitle

\IEEEdisplaynontitleabstractindextext

% \fi
\IEEEpeerreviewmaketitle

%\IEEEraisesectionheading{\section{Introduction}\label{sec:introduction}}

\section{Introduction}
\label{sec:intro}
Nonlinear shape deformation is ubiquitous in our every day life and simulating deformable objects has long been considered as an important yet challenging task for computer graphics and animation. In the past ten years, the finite element method (FEM) based frameworks~\cite{bathe2008finite} become more and more popular due to its versatility of encoding various material behaviors. With the prescribed external force $\mathbf{f}_\mathtt{ext}$, the dynamic equilibrium is forwarded by solving a high-dimensional nonlinear system of $\mathbf{f}(\mathbf{u})=\mathbf{f}_\mathtt{ext}$\footnote{Here $\mathbf{f}(\mathbf{u})$ is the general internal force, which consists of standard nonlinear internal force, damping force and inertial force. It is a function of the unknown displacement $\mathbf{u}$ after time derivative terms are linearized based on the chosen time integration method.} at each time step. Most nonlinear solvers start with an initial guess of the unknown displacement $\mathbf{u}$ and iteratively refine the result until the system converges in order to calculate the deformed model shape. While conceptually straightforward, the requirement of repetitive evaluations of the nonlinear internal force $\mathbf{f}_\mathtt{int}$ or/and its gradient $\partial\mathbf{f}_\mathtt{int}/\partial\mathbf{u}$ makes the simulation rather computational expensive.

Recently, the rapid development of the computing hardware
pushes forward the frontier of machine intelligence to an unprecedented
extend, and we have witnessed tremendous successes of utilizing carefully constructed deep neural networks (DNNs)~\cite{schmidhuber2015deep} in many classic computing problems like language processing~\cite{huang2013cross}, speech recognition~\cite{dahl2012context,pan2012investigation}, object tracking~\cite{wang2015visual,kristan2015visual} etc. With the support of sufficient ground truth data, DNN serves as a black box mapping its input to the output without the necessity of an explicit mathematical formulation. Since the FEM simulation is able to provide us as many as needed noise-free data, can we also exploit DNNs to deal with deformable simulation?

\begin{figure}[t!]
  \centering
  \includegraphics[width=\linewidth]{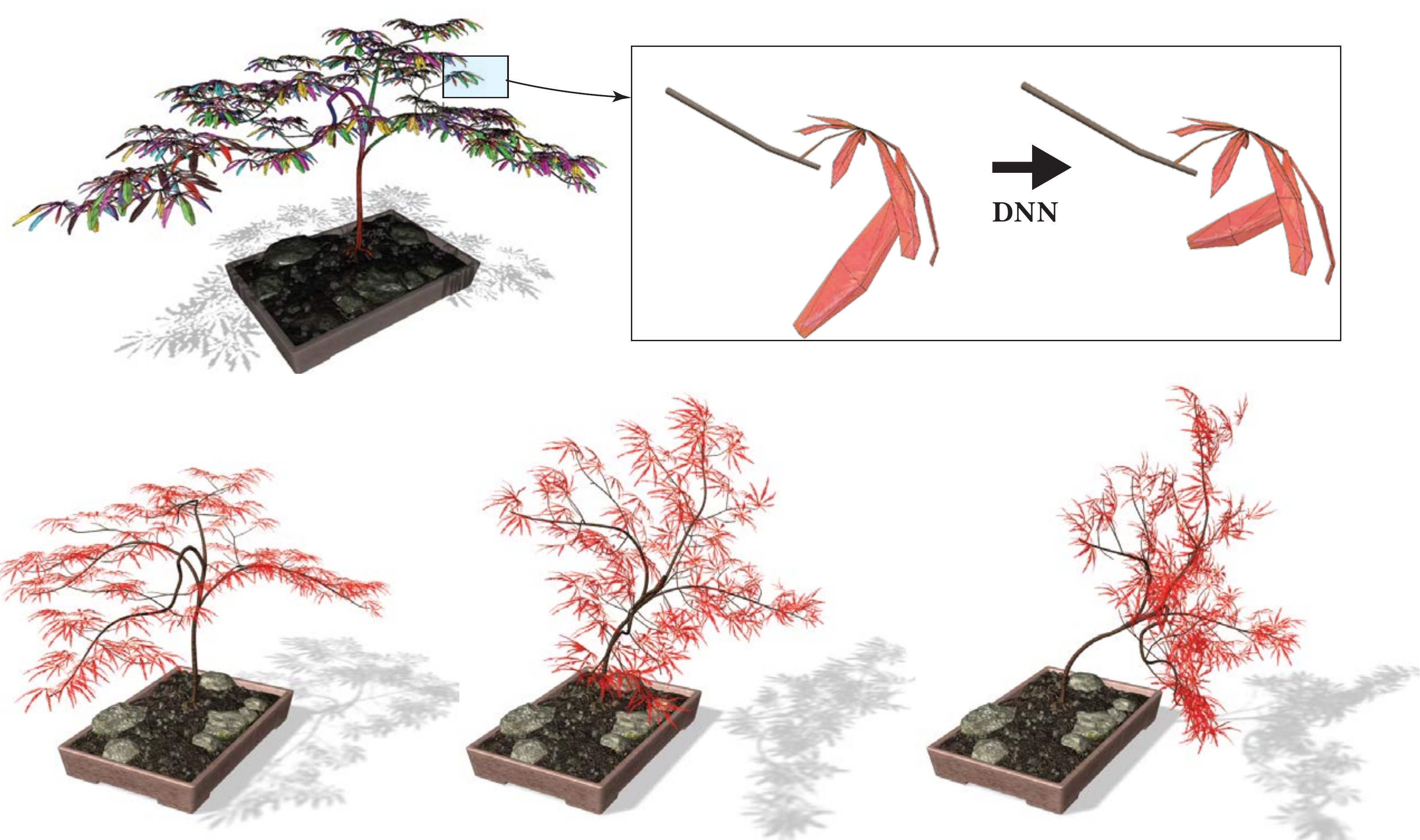}
  \caption{DeepWarp is a data-driven DNN-based nonlinear deformable simulator. By learning from full FEM simulation poses, it yields more accurate results than existing warping methods. We design three compact contextual features making the DNN training highly re-usable. In this example, the maple bonsai model consists of $255,552$ elements, and is decomposed into $1,771$ domains. A single DNN trained using a regular beam handles local dynamics for all the domains. High-quality animations with well-preserved local high-frequency deformations are produced at a near-interactive rate (5 FPS) without using model reduction.}\label{fig:teaser}
\end{figure}

At the first sight of the question, the answer seems to be positive because deformable simulation is essentially the reconstruction of the \emph{force-displacement relation} of an elastic body, and DNN is known to be skilled at expressing complex nonlinear relations~\cite{hornik1991approximation,cybenko1989approximation}. However, this problem is challenging in practice because the nonlinear force-displacement relation varies significantly (and intrinsically) under different simulation configurations such as domain geometry, discretization, boundary condition, constitutive law etc. If one chooses to build a network incorporating all the possible input permutations, the network would indubitably be an extremely huge and complex one. Even we manage to generate sufficient training data and optimize the network parameters to a reasonable level. A single forward pass of the network itself could take a longer time than running a regular FEM simulator due to the complexity of the network.
%Then, why bother using DNNs?
% -- after all, FEM itself has been serving as a standard tool for numerically approximate partial differential equations for over half century.
%\todo{simulation has a lot of clear data}\\

In this paper, we present a method, named \emph{DeepWarp}, to leverage DNNs to tackle intricate force-displacement relations of different nonlinear materials with a simple and light-weight network. As the name implies, our strategy is not to link the standard input ($\mathbf{f}_\mathtt{ext}$) and output ($\mathbf{u}$) of deformable simulation via a neural network directly. Instead, we map or \emph{warp} a simplified constitutive model $\mathscr{C}_0$ to a more complex and nonlinear one $\mathscr{C}_1$ using DNNs. It is expected that, the calculated displacement under $\mathscr{C}_0$ well encapsulates simulation configurations like force magnitude, domain tessellation and boundary conditions so that the remaining warp is local, and can be well fit by a simple net. To this end, we choose to use the linear elasticity for $\mathscr{C}_0$. The reasons are twofold. First, the linear elasticity has long been used to describe small-scale deformations (i.e. the infinitesimal strain theory). It is based on the Cauchy strain tensor, which is the first-order Taylor approximation of the full Green tensor. Second, as long as the deformable model is nonlinear, either material-wise or geometry-wise, the asymptotic complexity of simulating one time step is $\mathbf{O}(N^3)$ for the system of $N$ degrees of freedom (DOFs). Only the linear elasticity has $\mathbf{O}(N^2)$ complexity because of its constant system matrix. In other words, we gain a \emph{polynomial} performance profit by setting $\mathscr{C}_0$ as a linear model.

DeepWarp uses a single node-wise DNN to correct the nodal linear deformation to the corresponding nonlinear one. From this perspective, our method is conceptually similar to stiffness warping~\cite{muller2002stable} and modal warping~\cite{choi2005modal}, in which a linear solver is used after warping the deformed shape back to its undeformed orientation. We contrive three novel discriminative features as the input of the DNN, namely the \emph{geodesic}, \emph{potential} and \emph{digression}. We find that with these three descriptors, DeepWarp becomes fairly shape- and tessellation-independent, and the network trained with a simple model can be used to warp deformable bodies of distinctively different geometries making our DNN training highly re-usable. This important advantage is further enhanced when combined with the substructuring method~\cite{barbivc2011real}, where we decompose the input model into multiple convex domains, and run DeepWarp on each domain separately. All the examples reported in the paper except \figref{fig:t_bar} are based on the DNN trained using a regular rectangular beam model. DeepWarp is fast at both training stage and simulation stage. We utilize the rotation invariant feature of local deformation and compress the training set by at least an order. During the simulation run time, as DeepWarp only needs to perform a pre-factorized linear solve at each time step, it is able to handle large-scale models interactively.

%\todo{related work refine}\\
%\todo{technique part, convolutions}\\

\section{Related Work}
\label{sec:related_work}
The concept of deep learning can be dated back to late 1980s~\cite{dechter1986learning} in the machine learning community. Empowered by recent hardware advance, deep neural networks of various architectures have been harnessed to solve many long-standing computer vision problems such as recognition~\cite{sharif2014cnn,simonyan2014very}, classification~\cite{krizhevsky2012imagenet,farabet2013learning,karpathy2014large,he2015delving} and segmentation~\cite{chen2016deeplab,noh2015learning,girshick2016region}. Some existing methods are able to match or even beat human's vision perception system e.g. see the report from the ImageNet Large Scale Visual Recognition Challenge (ILSVRC)~\cite{ILSVRC15}. Given sufficient training data, DNNs provide a general ``template'' for the user to learn the input-output correspondence, which could be otherwise difficult or even impossible to be analytically formulated.

%An important reason that deep learning prevails in computer vision lies in the fact that images and videos can be naturally parameterized as an array of pixels. This allows a uniform and consistent input for a many vision problems. At the other end of the spectrum,

\vspace{5 pt}
\noindent\textbf{Learning for animation}\hspace{5 pt}
Indeed, the idea of learning is not new to computer animation, and it is also widely-known as the \emph{data driven} method~\cite{otaduy2012data}. For the \textbf{cloth animation}, low-resolution simulation can be enriched by using pre-computed high-resolution results with detailed wrinkles~\cite{wang2010example,kavan2011physics}. Wang et al.~\cite{wang2011data} built a piecewise linear stretching and bending model based on measured data to better depict the nonlinear dynamics of different cloth materials. Miguel et al.~\cite{miguel2012data} further enhanced this framework and recorded more deformation behaviors of the cloth simulation. Kim et al.~\cite{kim2013near} proposed a method to compress a large pre-simulation dataset so that these poses can be used at run time to improve the inertial cloth deformation. Following the similar idea, Xu et al.~\cite{xu2014sensitivity} blended pre-computed cloth shapes to directly synthesize the cloth deformation using the sensitivity analysis. Learning-based methods have also been popular for \textbf{motion and control} i.e. the reinforcement learning~\cite{coros2009robust,lee2010motion,peng2015dynamic,liu2016guided}. DNNs provide a convenient approach for further improving the learning effects~\cite{mnih2015human}. Following this direction, Liu et al.~\cite{liu2017learning} employed the deep Q-network to reorder existing control fragments and created necessary responses to unseen disturbances. Peng et al.~\cite{peng2017deeploco} used a DNN to train a high-level controller and a low-level one, which achieved robust locomotion coordinately. Holden et al.~\cite{holden2017phase} designed a phase-functioned neural network, whose weights are computed using a cyclic function. For \textbf{solid modeling}, learning is also a powerful tool, which allows the user to obtain actual physical parameters based on captured point cloud sequences~\cite{wang2015deformation}. Xu and Barbi\v{c}~\cite{Xu:2017:EDD:3072959.3073631} fine-tuned the damping model based on a few example deformations. Kim et al.~\cite{kim2017data} combined the physics-based simulation and data-driven to produce realistic soft tissue animation. Jones et al.~\cite{jones2016example} used the similar idea to simulate plastic deformation with a skinning-alike method. An et al.~\cite{an2008optimizing} proposed a learning-based numerical procedure named Cubature to efficiently evaluate the internal force and the force gradient during reduced deformable simulation. Deep learning also benefits the \textbf{fluid animation}. For instance, Ladicky et al.~\cite{Ladickydata2015} proposed a random forest based regression method to accelerate fluid simulation by predicting the kinematic configurations of particles based on a large training set. Chu and Thuerey~\cite{Chu:2017:DSS:3072959.3073643} used the convolutional neural networks (CNN) to extract necessary features to augment a coarse simulation and add back high-frequency details.

\vspace{5 pt}
\noindent\textbf{Nonlinear deformable simulation}\hspace{5 pt}
Physics-based deformable simulation has been an active research topic in graphics and animation since the exemplar work by Terzopoulos et al.~\cite{terzopoulos1987elastically}. While particle-based methods~\cite{muller2005meshless,pauly2005meshless,martin2010unified} or mass-spring systems~\cite{desbrun1999interactive,liu2013fast} are also legit, FEM becomes more widely-used~\cite{sifakis2012fem} for solid simulation. Wang et al.~\cite{wang2010multi} proposed a strain limiting method to increase the numerical stability for stiff deformable bodies. Alternatively, Irving et al.~\cite{irving2004invertible} tweaked the principle stress to resolve degenerated elements from extreme deformations. Forming the deformable simulation as a nonlinear optimization procedure, Hecht et al.~\cite{hecht2012updated} used an incremental Cholesky factorization scheme to lower the frequency of matrix re-factorization during the simulation. Zhu et al.~\cite{zhu2010efficient} adopted the multi-grid method to simulate high-resolution deformable volumes. Bouaziz et al.~\cite{bouaziz2014projective} introduced a robust local-global iterative solver named \emph{projective dynamics}. This idea later was generalized as the ADMM solver~\cite{Narain:2016:ASP:2982818.2982822} and synergized with Chebyshev ~\cite{wang2015chebyshev,wang2016descent}, L-BFGS~\cite{liu2017quasi} and GPU Gauss-Seidel~\cite{fratarcangeli2016vivace}. Accelerating nonlinear simulation can also be achieved by pre-computed models, for instance using modal analysis~\cite{Pentland:1989:GVM:74333.74355,barbivc2005real,yang2015expediting} or recent fullspace simulations~\cite{kim2009skipping}. Also known as model reduction methods, it is assumed that the deformed shape be a linear combination of those pre-computed poses or \emph{modes} so that the simulation can be projected into the spanned subspace. In an asymptotic sense however, model reduction is not better than regular simulation as the time complexity remains cubic w.r.t. the number of simulation DOFs.

\vspace{5 pt}
\noindent\textbf{Warping methods}\hspace{5 pt}
In this paper, we re-investigate this classic animation problem of nonlinear deformable simulation from a data-driven angle by shaping it as a nonlinear regression using the DNN. Unfortunately, the full spectrum of the force-displacement relation is complex and sensitive to the variance of  simulation settings. For instance, modifying the boundary condition (the anchor nodes of an FE mesh) could completely alter the deformed shape even with other simulation parameters unchanged. Besides, the dynamic simulation is essentially 4D -- the kinematic status of the deformable body does not only depend on its current external stimuli but also on its motion trajectory. To circumvent these two practical obstacles, we forge our regression based on the simulation result obtained using the linear elasticity. This idea is not new in graphics. An epic example would be the stiffness warping~\cite{muller2002stable}, which re-used the linear stiffness matrix by un-rotating the external force back to the model's rest shape orientation. Similarly, modal warping~\cite{choi2005modal} and rotation-strain coordinate~\cite{huang2011interactive,pan2015subspace} embedded a local coordinate frame at each node/element to relieve the artifacts of the linear elasticity under rotational deformation. This idea was also used for geometrically constructing nonlinear modes~\cite{von2013efficient}. These geometric warping techniques have been proven effective for animation editing~\cite{barbivc2012interactive,li2014space}, which requires performing high-dimensional space-time optimization.

Solving the linear elasticity encodes many simulation parameters such as boundary condition, tessellation resolution, external force etc. into the resulting linear displacement vector. On the top of this, we train a neural network to further correct the result to be a plausible and nonlinear one without worrying about accommodating all the simulation settings into the net.
%In addition, we use a semi-implicit integration to decouple inertial and internal forces so that each simulation time step can be linearized as a nonlinear equilibrium, which is then dealt with our DNN-augmented linear solver.
Our training is re-usable --
% -- \todo{both at the training stage and simulation stage}. Unlike existing DNN-facilitated simulation methods~\cite{Chu:2017:DSS:3072959.3073643}, which impose very large training sets to the network,
we train a DNN using a regular model of few thousand elements, and the network can then be used to handle a wide range of geometrically complex deformable bodies. During the simulation, because the system matrix for the linear elasticity is constant and pre-factorized, we obtain $\mathbf{O}(N^2)$ run-time complexity in fullspace, which is polynomially faster than existing nonlinear solvers.

\section{DNN Feature Vector}
\label{sec:feature}
The underlying mathematical relations between external forces and displacements of elastic bodies could be intrinsically changed under different simulation settings, and it is impossible in practice to encode the entire simulation configuration into a feature vector and feed to a neural network. Therefore, the primary challenge we are facing is to figure out an \emph{informative} and \emph{compact} feature vector as the DNN input. Informative refers to the discriminability of the feature so that an irrelevant training instance does not interfere. Compact means the feature should also be general so that the built network is small and light-weight. In this section, we start with a short review of the deformable model, pointing out that while the simulation is sophisticated, the linear-nonlinear deformation map of a small local volume is actually smooth. On the top of it, we show that our \emph{heuristic} feature vector augments the extracted kinematic information and produces plausible results.

%Unfortunately, these two criteria are contradict
%The most important step for
\subsection{{Deformable model: a quick review}}
Given an arbitrary material point $x$ on the deformable body, its deformation gradient $\mathbf{F}\in\mathbb{R}^{3\times3}$ is computed as $\mathbf{F}=\partial\mathbf{x}/\partial\bar{\mathbf{x}}$, where $\bar{\mathbf{x}}$ and $\mathbf{x}$ denote its rest shape position and the deformed position. Alternatively, we can also express $\mathbf{x}$ using its displacement $\mathbf{u}$ as $\mathbf{x}=\bar{\mathbf{x}}+\mathbf{u}$. Let $\mathbf{G}=\partial\mathbf{u}/\partial\bar{\mathbf{x}}$ and we name this 3 by 3 tensor as \emph{displacement gradient tensor}. It is easy to verify that $\mathbf{F}=\mathbf{G}+\mathbf{I}$. Under the linear elasticity, the deformation is described using the Cauchy strain: $\widetilde{\epsilon}=\frac{1}{2}(\mathbf{G}+\mathbf{G}^\top)$, and the strain energy density $\widetilde{\Psi}$ is:
\begin{equation}\label{eq:linear_strain_energy}
\widetilde{\Psi}=\frac{k}{2(1+\nu)}\widetilde{\epsilon}:\widetilde{\epsilon}+\frac{k\nu}{2(1+\nu)(1-2\nu)}\mathtt{tr}^2(\widetilde{\epsilon}).
\end{equation}
Here $k$ and $\nu$ are the Young's modulus and Poisson's ratio. Clearly, $\widetilde{\Psi}$ is a quadratic function of $\mathbf{G}$. Therefore, the corresponding Piola stress becomes a linear function of $\mathbf{G}$:
\begin{equation}\label{eq:linear_stress}
\widetilde{\mathbf{P}}=\frac{k}{2(1+\nu)}\big(\mathbf{G}+\mathbf{G}^\top\big)+\frac{k\nu\cdot\mathtt{tr}(\mathbf{G})}{(1+\nu)(1-2\nu)}\mathbf{I}.
\end{equation}
For most other hyperelastic materials, the deformation is actually described with the Green strain: $\epsilon=\frac{1}{2}(\mathbf{F}\mathbf{F}^\top-\mathbf{I})=\widetilde{\epsilon}+\frac{1}{2}\mathbf{G}\mathbf{G}^\top$. One can see that the Cauchy strain used in linear elasticity is simply the linear portion of the Green strain. Take the St. Venant-Kirchhoff (StVK) material an example, whose strain energy density is formulated by replacing $\widetilde{\epsilon}$ by $\epsilon$:
\begin{equation}\label{eq:stvk_strain_energy}
\Psi_\mathtt{StVK}=\frac{k}{2(1+\nu)}\epsilon:\epsilon+\frac{k\nu}{2(1+\nu)(1-2\nu)}\mathtt{tr}^2(\epsilon),
\end{equation}
which is a fourth-order polynomial of the displacement gradient $\mathbf{G}$, and its stress is cubically related to $\mathbf{G}$. With the help of FEM, the differential strain-stress relation is integrated and becomes the macroscopic force-displacement relation that we are interested in.

\setlength{\columnsep}{5 pt}
\begin{wrapfigure}[8]{r}{0.55\linewidth}
\vspace{-5 pt}
%\hspace{-1.35cm}
%\begin{center}
\includegraphics[width =\linewidth]{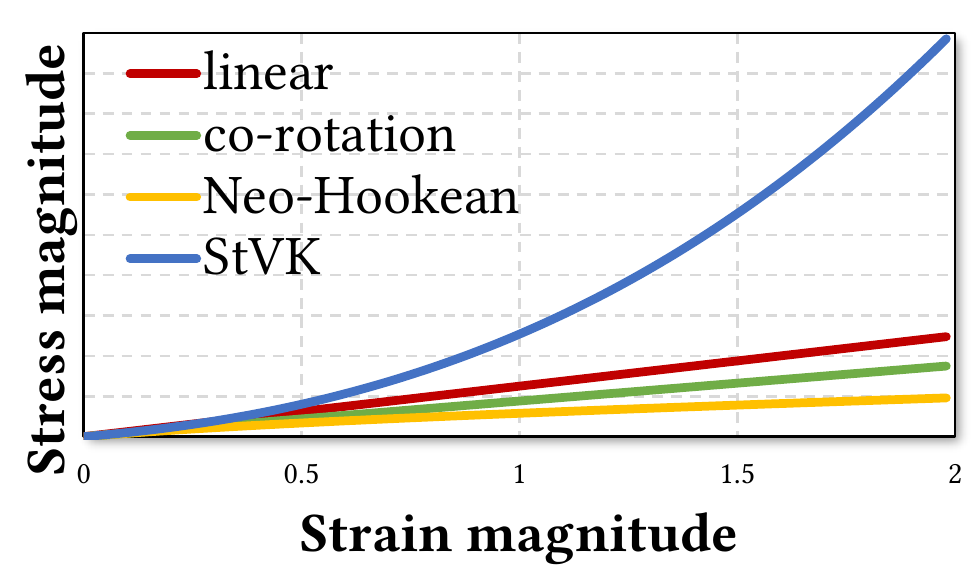}
\vspace{-20 pt}
\end{wrapfigure}
In reality, the \emph{magnitude} of a deformation, which may be imprecisely understood as $|\mathbf{G}|$, is typically small. For instance, doubling the length of an elastic rope by stretching is considered a very large deformation where $|\mathbf{G}|=1$. Besides, strain-stress curves for various materials are all aligned at the origin (a zero strain yields a zero stress) and within the same monotonically increasing interval (a larger strain yields a larger stress in general). This implies that even though a linear function could deviate a lot from a cubic one (or other nonlinear relations), the strain-stress curves of the linear elasticity and a nonlinear elasticity do not fundamentally differ from each other in regular deformable simulations. An example is given in the inset figure, where we plot the strain-stress curves of the linear, co-rotation, StVK and Neo-Hookean laws under a rotation-free linear stretch.

\vspace{5 pt}
\noindent\textbf{Geometric warp}\hspace{5 pt}
In fact, the dominant factor drives the linear elasticity away from a nonlinear counterpart is not the material nonlinearity, but the geometry nonlinearity. This is because a rigid, deformation-free rotation leads to a non-zero Cauchy strain, which produces unrealistic deformation effects. Under this consideration, the modal warping (MW) technique~\cite{choi2005modal} embeds each node on the mesh a local frame. The curl of local displacement field around the $i$-th node is calculated: $\mathbf{w}_i=\nabla\times\mathbf{u}_i$. If it takes a unit time to displace node $i$ from $\bar{\mathbf{x}}_i$ to $\bar{\mathbf{x}}_i+\mathbf{u}_i$, $\mathbf{u}_i$ also represents its velocity at $t=1$. $\mathbf{w}_i$ can then be understood as its angular velocity at the same moment. Based on this assumption, MW linearly ramps the angular velocity from the rest shape to the current time instance $t$ and calculates a warp transformation as:
\begin{equation}\label{eq:mw}
\mathbf{W}_\mathtt{MW}=\frac{1}{t}\int_0^t\mathtt{exp}\left(\frac{\tau}{t}[\mathbf{w}_i]_\times\right)\mathrm{d}\tau,
\end{equation}
where $[\mathbf{w}_i]_\times$ is the skew symmetric matrix of $\mathbf{w}_i$. Similarly, one can use rotation-strain coordinate by decomposing the $\mathbf{G}_i$ into a skew symmetric part:  $[\mathbf{w}_i]_\times=(\mathbf{G}_i-\mathbf{G}_i^\top)/2$ and a symmetric part: $\mathbf{S}_i=(\mathbf{G}_i+\mathbf{G}_i^\top)/2$~\cite{huang2011interactive,pan2015subspace}. The rotation-strain warp (RSW) transformation can then be computed treating $\mathbf{w}_i$ as an Euler vector:
\begin{equation}\label{eq:rsw}
\mathbf{W}_\mathtt{RSW}=\mathtt{exp}\left([\mathbf{w}_i]_\times\right)(\mathbf{S}_i+\mathbf{I})-\mathbf{I}.
\end{equation}
While not physically accurate, these geometric warping methods produce visually pleasing shapes and have been used in many time-critical graphics applications~\cite{li2014space,barbivc2012interactive}.

\subsection{Linear-nonlinear correspondence}
DeepWarp is inspired by the encouraging results from the existing warping methods. However, DeepWarp does not explicitly assume a fixed nonlinear regression formula as Eqs.~\eqref{eq:mw}~or~\eqref{eq:rsw} do. Instead, we train a DNN to obtain a more accurate regression based on full simulations. The key question here is how to determine what is the ``right'' deformation that corresponds to the one calculated using the linear elasticity.

\begin{figure}[h!]
  \centering
  \includegraphics[width=0.9\linewidth]{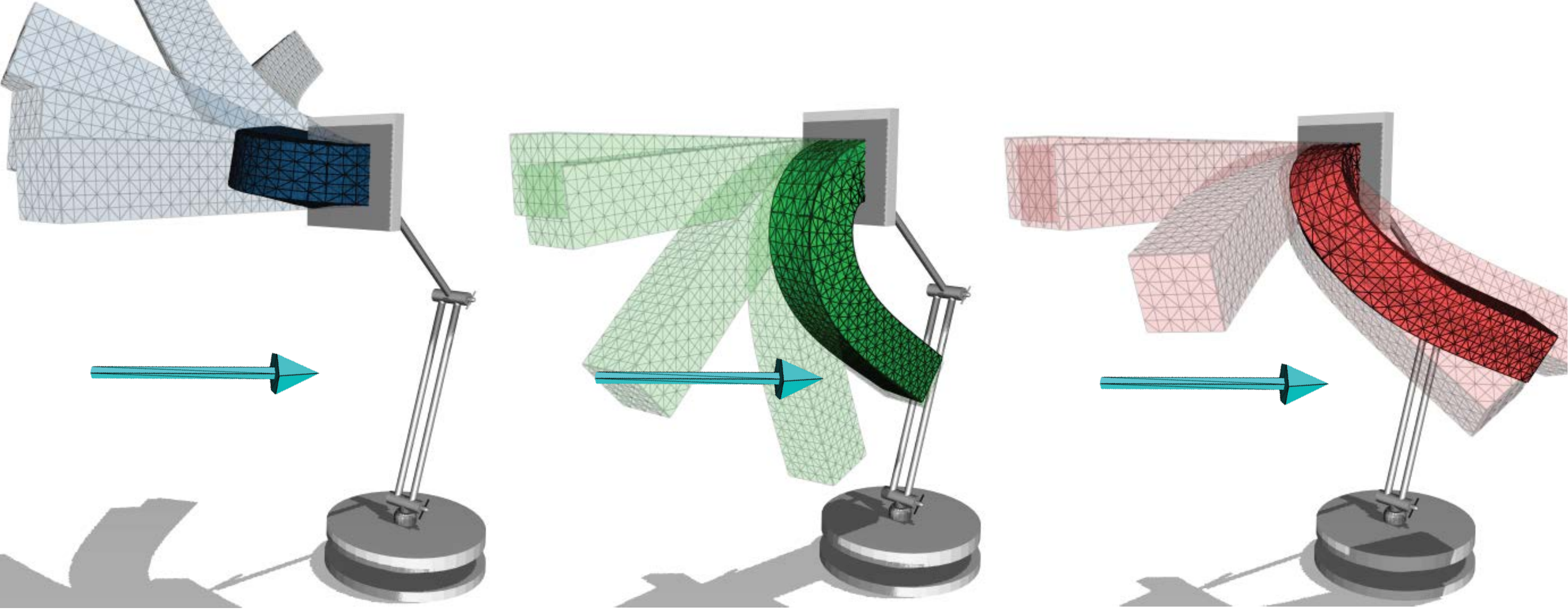}
  \caption{Different motion trajectories lead to different equilibrium shapes even under the same external force (highlighted as the blue arrow).}\label{fig:local_min}
\end{figure}

A na\"ive thought is to solve the quasi-static equilibrium of $\mathbf{f}_\mathtt{int}(\mathbf{u})=\mathbf{f}_\mathtt{ext}$ for a deformable body under the same external force and boundary condition using the linear elasticity and a nonlinear constitutive model. Unfortunately, this method is only valid for small deformations. Under large external forces, the nonlinear system could reach multiple local minima with different shapes, and which one being reached is context-dependent i.e. up to the history of the deformation trajectory (\figref{fig:local_min}). From a numerical point of view, the solution of $\mathbf{f}_\mathtt{int}(\mathbf{u})=\mathbf{f}_\mathtt{ext}$ depends on the initial guess of $\mathbf{u}$ and the strategy of computing $\Delta\mathbf{u}$ during the iteration. The iteration may not converge to the global minimum if the starting guess is far away from it.

Our solution to this problem is to register a linear deformation \emph{sequence} to a nonlinear one starting from the rest shape. Specifically, given an external force $\mathbf{f}_\mathtt{ext}$, we compute a series of quasi-static linear deformation by solving the Euler-Lagrange equation with an increased mass damping so that the acceleration at each time step is negligible. Each time step yields a linear displacement vector $\widetilde{\mathbf{u}}$, and we estimate a local rotation for the $i$-th node as:
\begin{equation}\label{eq:local_rotation}
\mathbf{R}_i=\mathtt{exp}\big(\left[\nabla\times\big(\mathbf{P}_i\mathbf{P}^\top_i\big)^{-1}\mathbf{P}^\top_i\mathbf{U}_i\right]_\times\big),
\end{equation}
where columns in $\mathbf{P}_i$ and $\mathbf{U}_i$ are rest shape positions and displacements of neighbor nodes adjacent to $i$. In other words, $\big(\mathbf{P}_i\mathbf{P}_i^\top\big)^{-1}\mathbf{P}^\top_i\mathbf{U}_i$ gives a least-square evaluation of $\mathbf{G}$ around the $i$-th node. The linear internal force at the current time step is $\widetilde{\mathbf{f}}_\mathtt{int}=\mathbf{K}\widetilde{\mathbf{u}}$. Note that $\widetilde{\mathbf{f}}_\mathtt{int}\neq\mathbf{f}_\mathtt{exp}$ until the final equilibrium is reached due to the existence of the damping. Afterwards, the corresponding nonlinear deformation $\mathbf{u}$ is obtained by solving:
\begin{equation}\label{eq:nonlinear_displacement}
\min_{\mathbf{u}} |\mathbf{f}_\mathtt{int}(\mathbf{u})-\mathscr{R}\mathbf{K}\widetilde{\mathbf{u}}|,
\end{equation}
where $\mathscr{R}$ is a block-diagonal matrix, and each of its 3 by 3 diagonal block is the estimated nodal rotation computed via \eqnref{eq:local_rotation}. We use Newton's method to solve \eqnref{eq:nonlinear_displacement} by setting the initial guess of $\mathbf{u}$ as the solution in the previous time step. In our implementation, we notice that Newton's method occasionally fails during the iteration. Therefore, we impose the Wolfe condition~\cite{wolfe1969convergence} to adjust the step length.
\begin{figure*}[t!]
  \centering
  \includegraphics[width=0.9\linewidth]{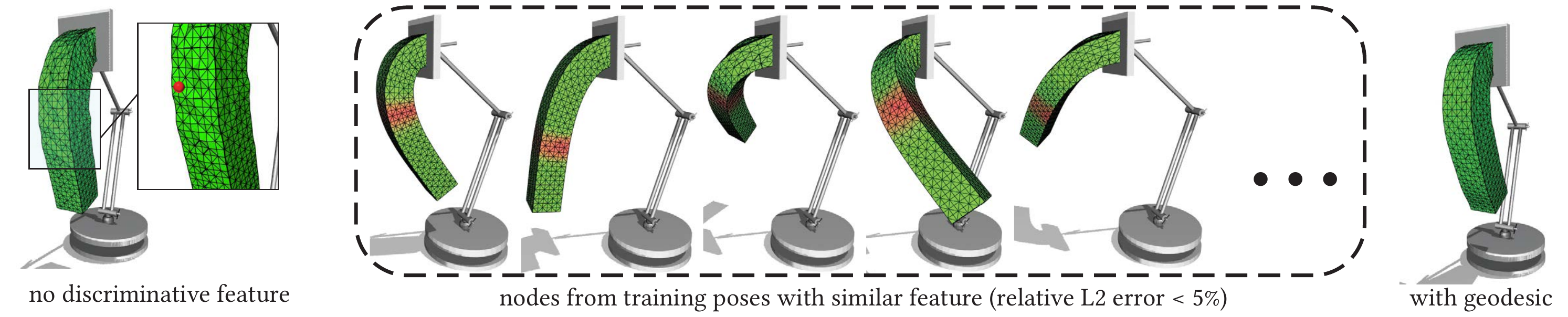}
  \caption{Only using kinematic feature as the input of DNN yields noticeable jittery artifacts. A node, because of its kinematic feature is not discriminative, could be influenced by many irrelevant instances in the training date. Large discrepancies among these mis-matched nodes induce high-frequency variations of the DNN mapping. Incorporating geodesic feature effectively eliminates this artifact.}\label{fig:geodesic}
\end{figure*}

In our DNN training, we simplify the external force setting by only considering two types of $\mathbf{f}_\mathtt{ext}$: directional force field and circular force field. The directional field uses a prescribed force direction, while the force direction in the circular field follows the tangent direction of a set of concentric circles. Such simplification frees us from generating an overwhelmingly large training set due to diverse external force conditions. Its limitation is also obvious: DeepWarp may lose some local deformation effects induced by high-frequency external force.
%Solving the linear system also encodes many simulation settings that are not intuitive to be parameterized such as the tessellation, external forces, boundary conditions etc.

\subsection{Discriminative feature}
With paired $\langle\widetilde{\mathbf{u}},\mathbf{u}\rangle$, we can build a node-wise regression machine using a DNN that replaces \eqnref{eq:mw} or \eqnref{eq:rsw}. For the $i$-th node, in addition to its linear displacement $\widetilde{\mathbf{u}}_i$, the rotation information of its local displacement gradient $\mathbf{G}_i$ is directly pertinent to the warp transformation, and should be passed to the DNN as the input. To this end, we choose to use the skew symmetric part of $\mathbf{G}_i$ and represent it as a 3-vector as in~\cite{huang2011interactive}. However, only feeding these two pieces of information to the DNN is not enough, and the resulting deformation appears jittery and non-smooth as shown in \figref{fig:geodesic}. In this example, we use the Neo-Hookean elasticity, whose strain energy is:
\begin{equation}\label{eq:neohookean_energy}
\Psi_\mathtt{NH}=\frac{k}{4(1+\nu)}[I_1-\log(I_3)-3]+\frac{k\nu}{8(1+\nu)(1-2\nu)}\log^2(I_3).
\end{equation}
$I_1$ and $I_2$ are the invariants of the deformation gradient, defined based on $\mathbf{F}$'s singular values $\sigma_1$, $\sigma_2$ and $\sigma_3$ such that: $I_1=\sigma^2_1+\sigma^2_2+\sigma^2_3$
and $I_3=\sigma^2_1\sigma^2_2\sigma^2_3$.

This artifact was also noticed and discussed in previous data-driven simulation literature~\cite{Ladickydata2015}, which is because pure node-wise kinematic features do not contain sufficient contextual information, and thus are not discriminative to reach a conclusive per-node linear-nonlinear map. To further illustrate this artifact, we pick a jittery node (marked as a red sphere in the figure) and inversely query for nodes in our training set that have similar features ($<5\%$ relative L2 error w.r.t. the feature vector from the picked node). We can see from the figure that there are a number of training poses having multiple nodes (on the red areas) with very similar feature vectors as the input. In other words, the final displacement of the picked node becomes a certain mixture of displacements of many distant and irrelevant nodes. Such ambiguity of pure kinematic feature is the primary reason behind this artifact.

  %\todo{Please refer to \secref{sec:experiment} for details about the training data generation.}

One of our contribution is to design a compact contextual feature to resolve this mis-math. While one could follow the method used in~\cite{Ladickydata2015} to use the integral features of local dynamic parameters around a node, we found that our simple strategy yields satisfying result. We speculate that this is because DOFs in solid simulation are more tightly coupled than in fluid simulation~\cite{Ladickydata2015}. An important advantage of such compactness is that the DNN training is also quite fast. Compared with state-of-the-art pre-computed deformable models i.e.~\cite{an2008optimizing}, we can finish training in a few minutes, and the obtained DNN can be applied to a wide range of models.

\vspace{5 pt}
\noindent\textbf{Discriminative feature~\RNum{1}: geodesic}\hspace{5 pt}
The geodesic of a node $i$, $g_i$ is the normalized length of the shortest path from node $i$ to its nearest anchor node within the deformable body. For a training model, we first uniformly scale it to fit a unit bounding sphere. Then, we compute the shortest path using the Dijkstra's algorithm for all the un-anchored nodes.
%In practice, anchor nodes are often picked interactively by the user i.e. via mouse dragging, which form a small anchor node cluster. In that case, we choose the anchor node that is closest to the geometry center of the selected anchor cluster as the source for the Dijkstra's algorithm.
Lastly, calculated path lengths are scaled by the maximum geodesic so that all the $g$ values are within the normalized interval of $[0,1]$. Our heuristic of choosing the geodesic feature is based on the observation that if a node is closer to an anchor node, it trends to have less deformation than nodes that are away from it. By inducing the geodesic feature, a node far from anchor nodes does not miss-pair to a node close to anchor nodes only because the it undertakes a smaller external force. Consequently, the jittery artifact is effectively removed as shown in the rightmost snapshot in \figref{fig:geodesic}.

\begin{figure}[h!]
  \centering
  \includegraphics[width=0.9\linewidth]{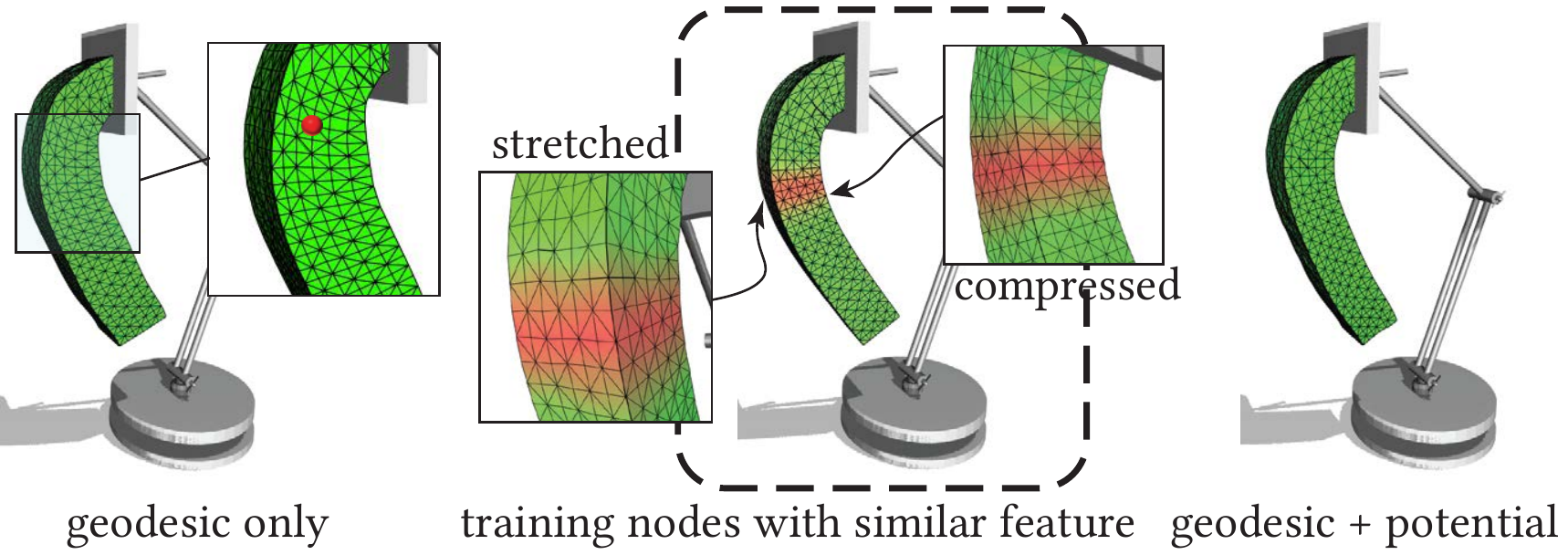}
  \caption{Volume expansion artifact remains even with the geodesic feature added. This is because the nodes with similar geodesic value may have different internal tractions. We use the potential feature to sort the training data to avoid this issue.}\label{fig:potential}
\end{figure}
\begin{figure*}[t!]
  \centering
  \includegraphics[width=\linewidth]{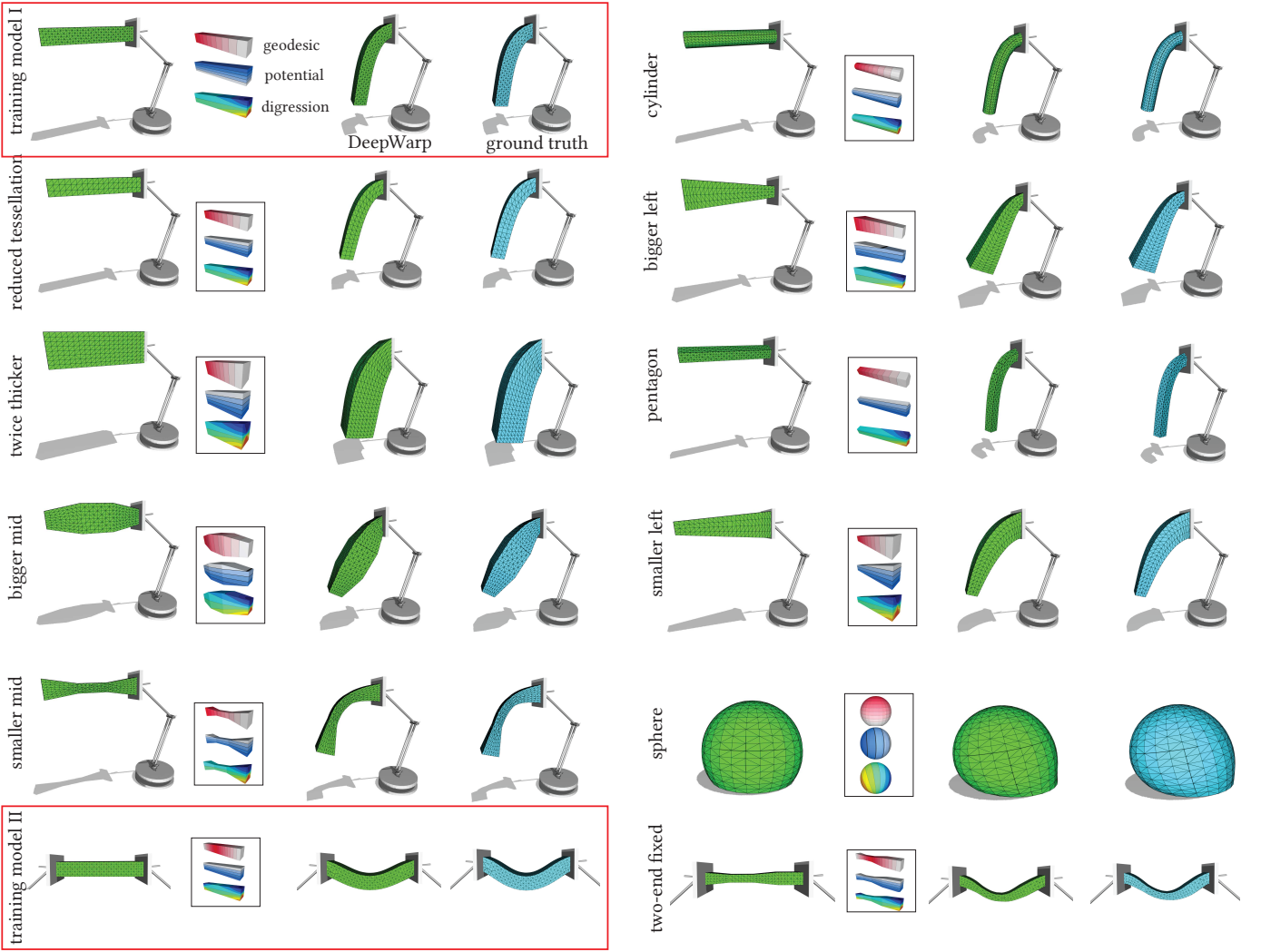}
  \caption{We test the generality of the designed features on a variety of shapes. The training data are generated using the standard rectangular beam (highlighted by a red box). The resulting DNN successfully handles many beam-like models but with distinctively different shapes. The distributions of three features are also plotted.}\label{fig:multi_shape}
\end{figure*}

\vspace{5 pt}
\noindent\textbf{Discriminative feature~\RNum{2}: potential}\hspace{5 pt}
Including the geodesic feature however, does not avoid the artifact of volume increase and shrinkage. As shown in \figref{fig:potential}, bending the beam also increases its volume noticeably, especially at curved areas. In order to dig out the missing contextual information behind this issue, we use the similar approach by picking a node within the problematic area and query the instances from our training set that have a similar feature of the selected one. We can see from the figure that, thanks to the incorporated geodesic feature, now this selected node only pairs with a training pose under a very similar deformation. However, it still matches multiple nodes on this pose. This is because the beam is a symmetric shape, and a loop of nodes on its surface have similar geodesic values -- among which, some are stretched and some are compressed. Without being able to distinguish these contexts, the volume of the warped model is likely to shrink or expand unnaturally.

\setlength{\columnsep}{5 pt}
\begin{wrapfigure}{r}{0.46\linewidth}
\vspace{-20 pt}
%\hspace{-1.35cm}
%\begin{center}
\includegraphics[width =\linewidth]{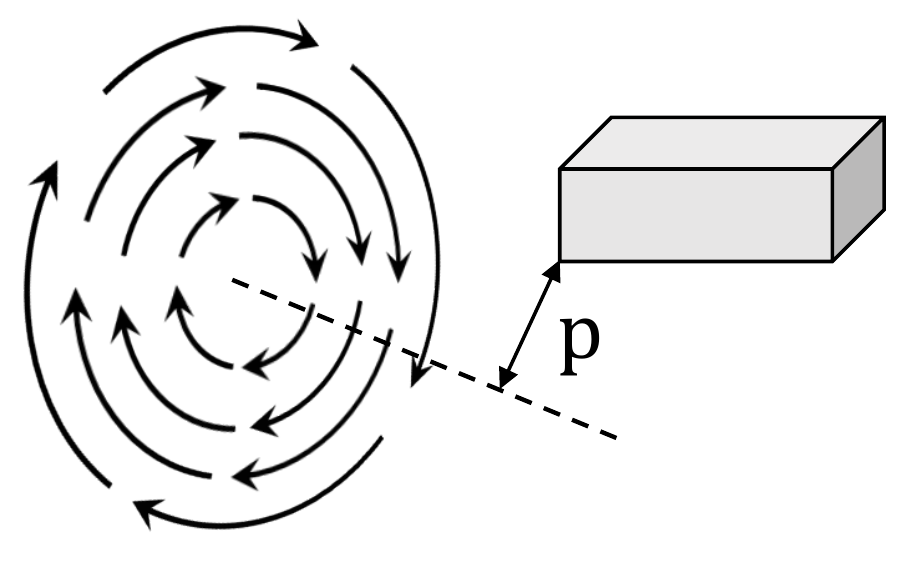}
\vspace{-20 pt}
\end{wrapfigure}
We notice that whether nodes are being stretched or compressed typically depends on their relative positions in the applied force field. Therefore, we introduce another scaler feature named potential $p$ to resolve this ambiguity. If a directional force field is applied, for each node on the mesh, we project its rest shape position onto the force direction and re-map the resulting projections to the interval of $[0,1]$. On the other hand, if a circular force field is applied, the potential of a node is the distance between its rest shape position and the circular axis, as shown in the inset figure. This value is also scaled to $[0,1]$. As we can see from~\figref{fig:potential}, the deformation of the beam model is almost identical to the one obtained using the full simulation after we inject the potential feature into the DNN.

\vspace{5 pt}
\noindent\textbf{Discriminative feature~\RNum{3}: digression}\hspace{5 pt}
So far, we generate a set of registered linear and nonlinear poses of the beam model. Node-wise linear to nonlinear deviation is learnt by a DNN, which is then used to warp a linear displacement of the same model to obtain its nonlinear shape. While the results are visually plausible, real-world applications will require deformable animations of various 3D models. DeepWarp becomes cumbersome and less practical if one needs to re-train a DNN for each different deformable body.
\begin{figure}[th!]
  \centering
  \includegraphics[width=\linewidth]{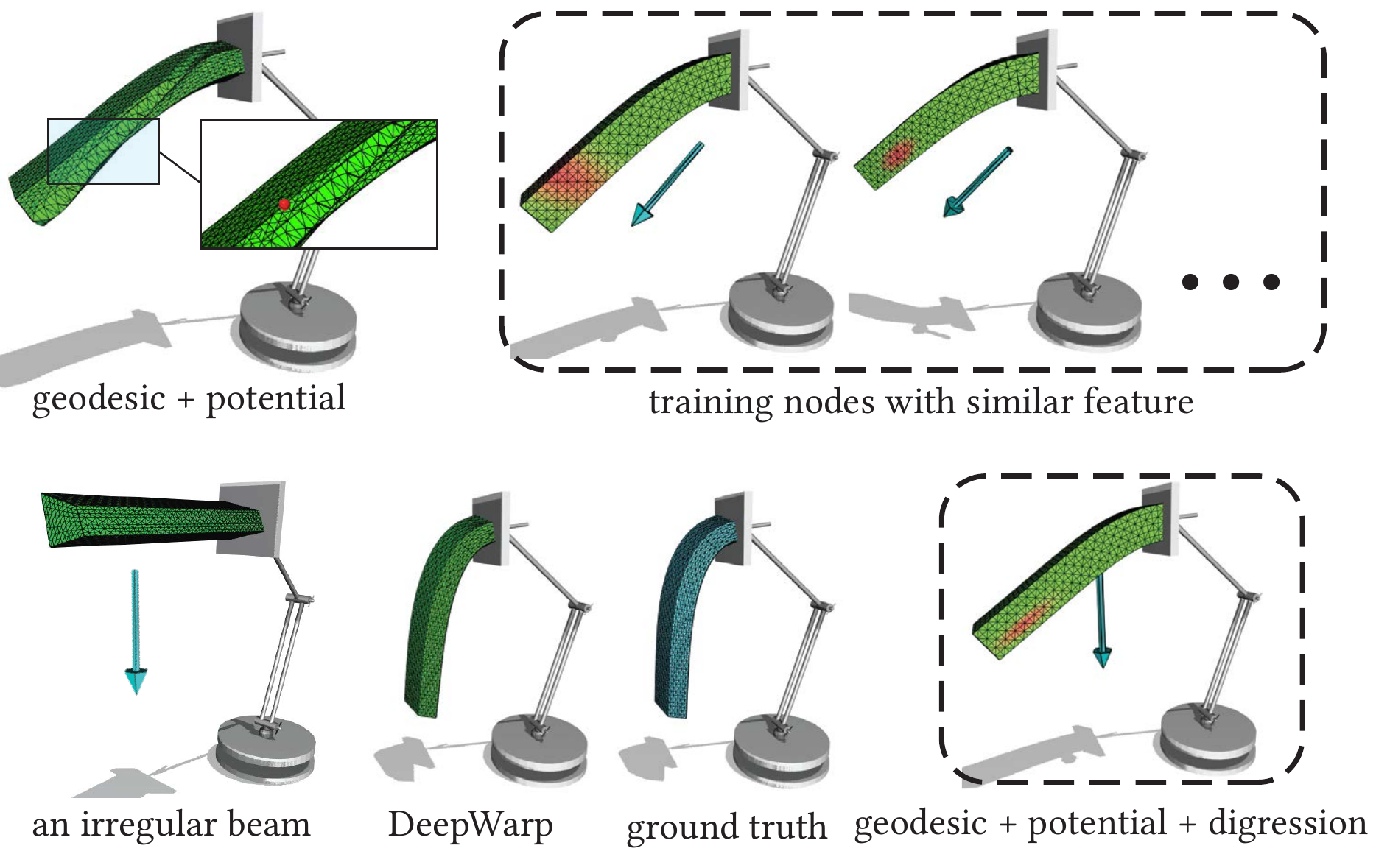}
  \caption{In order to make DeepWarp re-usable for various deformable bodies, we use the digression as our third discriminative feature. With this feature included, DNN is able to handle an irregular beam based on the training set generated using a standard rectangular beam model.}\label{fig:digression}
\end{figure}

Unfortunately, if we alter the rest shape geometry of the beam model as shown in \figref{fig:digression}, unrealistic jittery deformations are observed again even after incorporating geodesic and potential features. By querying the training set, we see that because the updated shape is irregular and asymmetric, the most similar training poses become the ones under oblique force fields regardless an upright gravity force field is applied in the simulation. To further correct this mis-match, we use the digression feature $d$ to describe the nodal position w.r.t. the direction of the external force. Specifically, the digression for node $i$ is defined as:
\begin{equation}\label{eq:digression}
d_i =\arccos\left(\frac{\bar{\mathbf{x}}_i-\bar{\mathbf{x}}_a}{|\bar{\mathbf{x}}_i-\bar{\mathbf{x}}_a|}\cdot\frac{\mathbf{f}_\mathtt{ext}}{|\mathbf{f}_\mathtt{ext}|}\right),
\end{equation}
where $\bar{\mathbf{x}}_a$ is the rest shape position of the anchor node that is closest to node $i$. Indeed, digression sorts nodes based on their local orientational deviations from the external force direction. The digression feature ranges from $0$ to $\pi$. If a circular force field is applied, the digression is simply set as $-1$. As shown in \figref{fig:digression}, with geodesic, potential and digression included, the training data generated using a rectangular beam model can also be used to warp the irregular beam, and DeepWarp produces high-quality nonlinear deformation.

\begin{figure}[h!]
  \centering
  \includegraphics[width=0.9\linewidth]{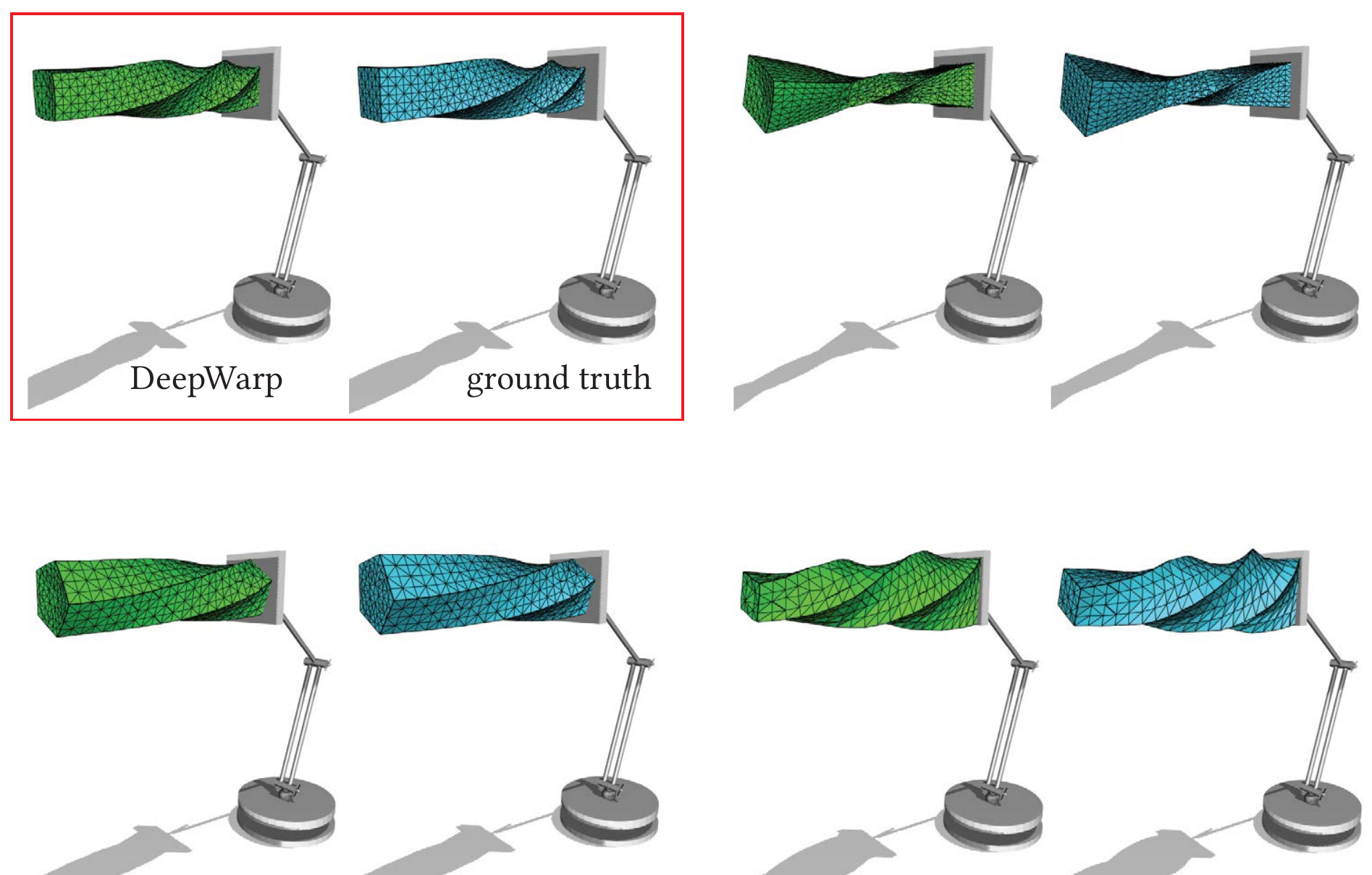}
  \caption{DeepWarp works well under circular force field too. In this case, the digression feature is set as -1 for all the nodes on the mesh.}\label{fig:twist}
\end{figure}
\begin{figure}[h!]
  \centering
  \includegraphics[width=\linewidth]{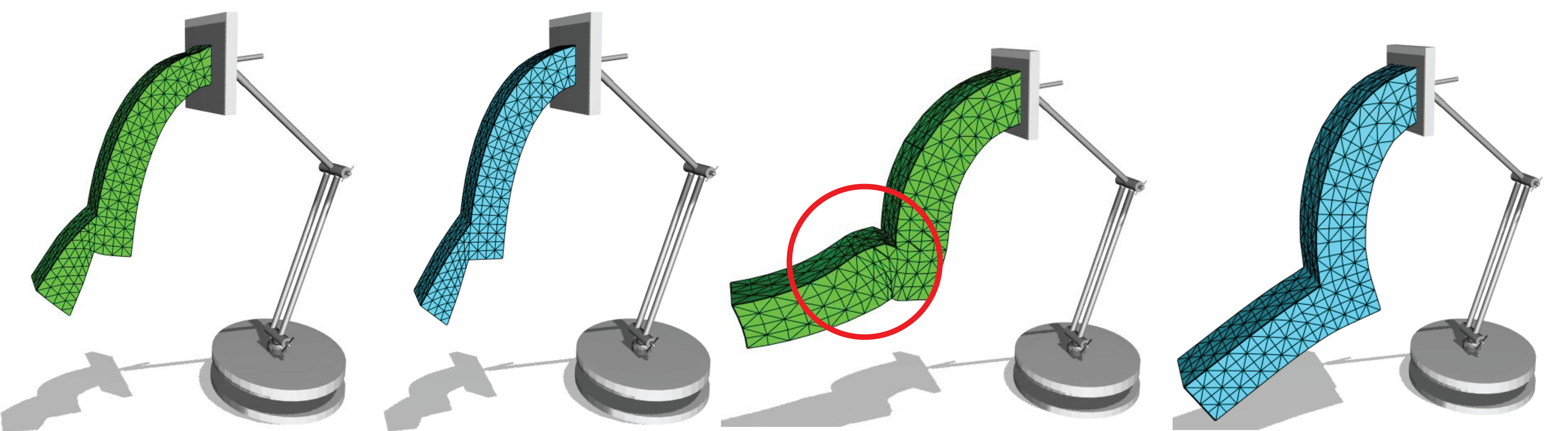}
  \caption{When the shape becomes more concave, DNN trained using a rectangular beam produces artifacts.}\label{fig:concave_fail}
\end{figure}

\vspace{5 pt}
\noindent\textbf{Discussion}\hspace{5 pt}
Our features allow the resulting DNN well handles models with various shapes, different tessellations and altered boundary conditions. More results can be found in Figs~\ref{fig:multi_shape}~and~\ref{fig:twist}. From these examples, readers may probably notice that geodesic, potential and digression features actually provide a volumetric \emph{parametrization} of deformable bodies so that models of different geometries and tessellations are somehow registered in a meaningful way, and node-wise DNN can then be applied. In fact, there are many elegant algorithms in graphics and computational geometry that generate the volumetric map between different shapes~\cite{xu2013biharmonic,fu2015computing,aigerman2013injective}. However in the context of DeepWarp, this volumetric map depends on the configuration of external force and boundary conditions. While existing methods may also be modified to incorporate these additional conditions or constraints, we found that our simple strategy suffices in most cases. An exception is reported in \figref{fig:concave_fail}, and we find that DeepWarp using a convex training model often fails when the deformable body gets more concave. Next, we will show how to walk around this limitation without re-training a DNN based on the target shape.
\label{sec:shape}

\section{Incorporate Complex Shapes}
\setlength{\columnsep}{5 pt}
\begin{wrapfigure}{r}{0.5\linewidth}
\vspace{-15 pt}
%\hspace{-1.35cm}
%\begin{center}
\includegraphics[width =\linewidth]{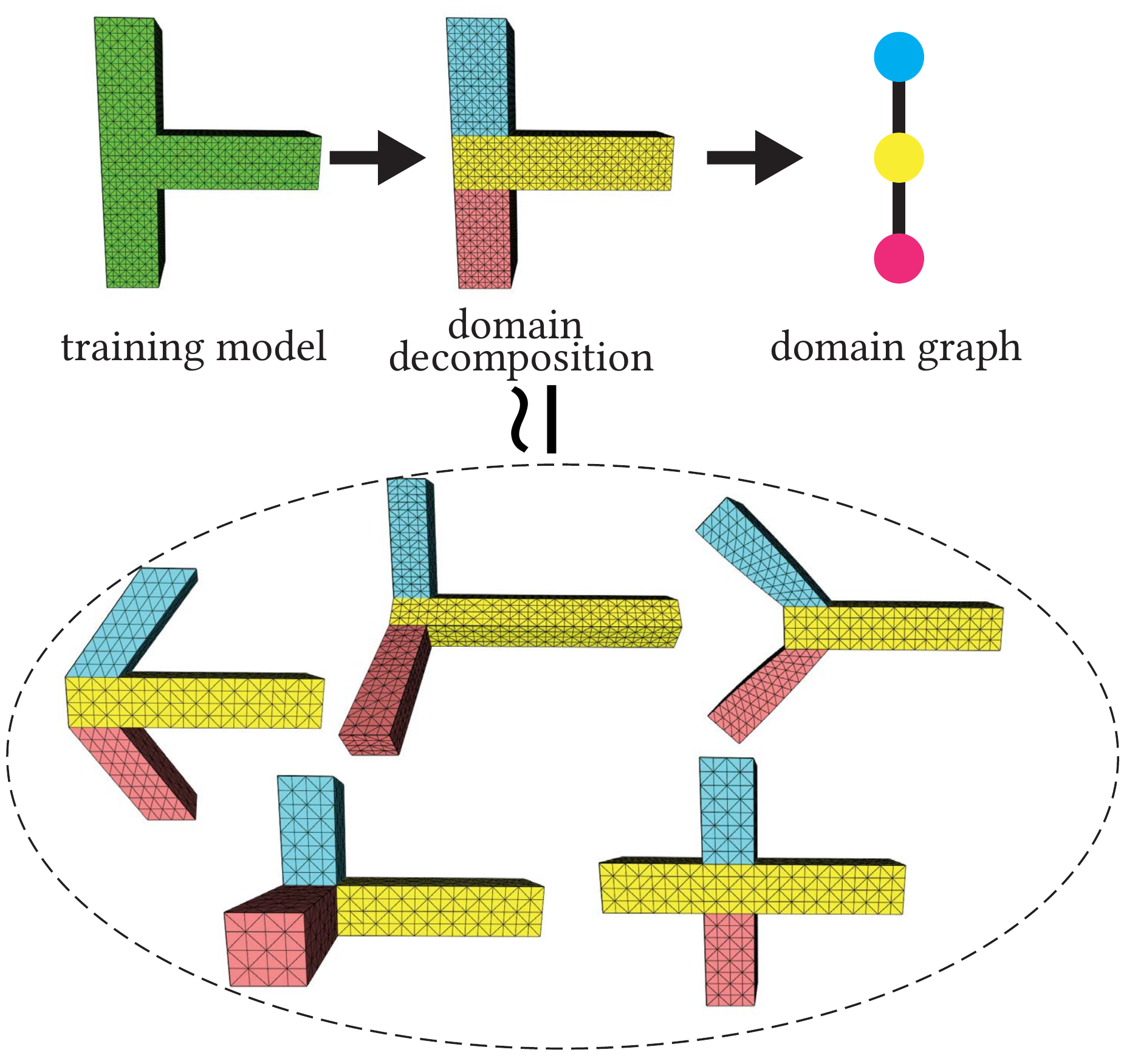}
\caption{Building domain graph for the domain decomposed model is an easy and effective way to identify shapes with similar concavity.}
\label{fig:concave}
\vspace{-5 pt}
\end{wrapfigure}
The exhaustiveness of 3D geometric diversity is endless. Obviously, training set generated with a single rectangular beam cannot cover all the different feature combinations. We find that the DNN trained using the beam model is able to deal with many convex 3D shapes (i.e. see \figref{fig:multi_shape}). However, it often fails when the target deformable body becomes more concave (\figref{fig:concave_fail}). A straightforward idea is to train the DNN using a model with similar concavity of the target deformable body, but \emph{how to describe the similarity of concavity among 3D shapes?}

\begin{figure*}[ht!]
  \centering
  \includegraphics[width=0.9\linewidth]{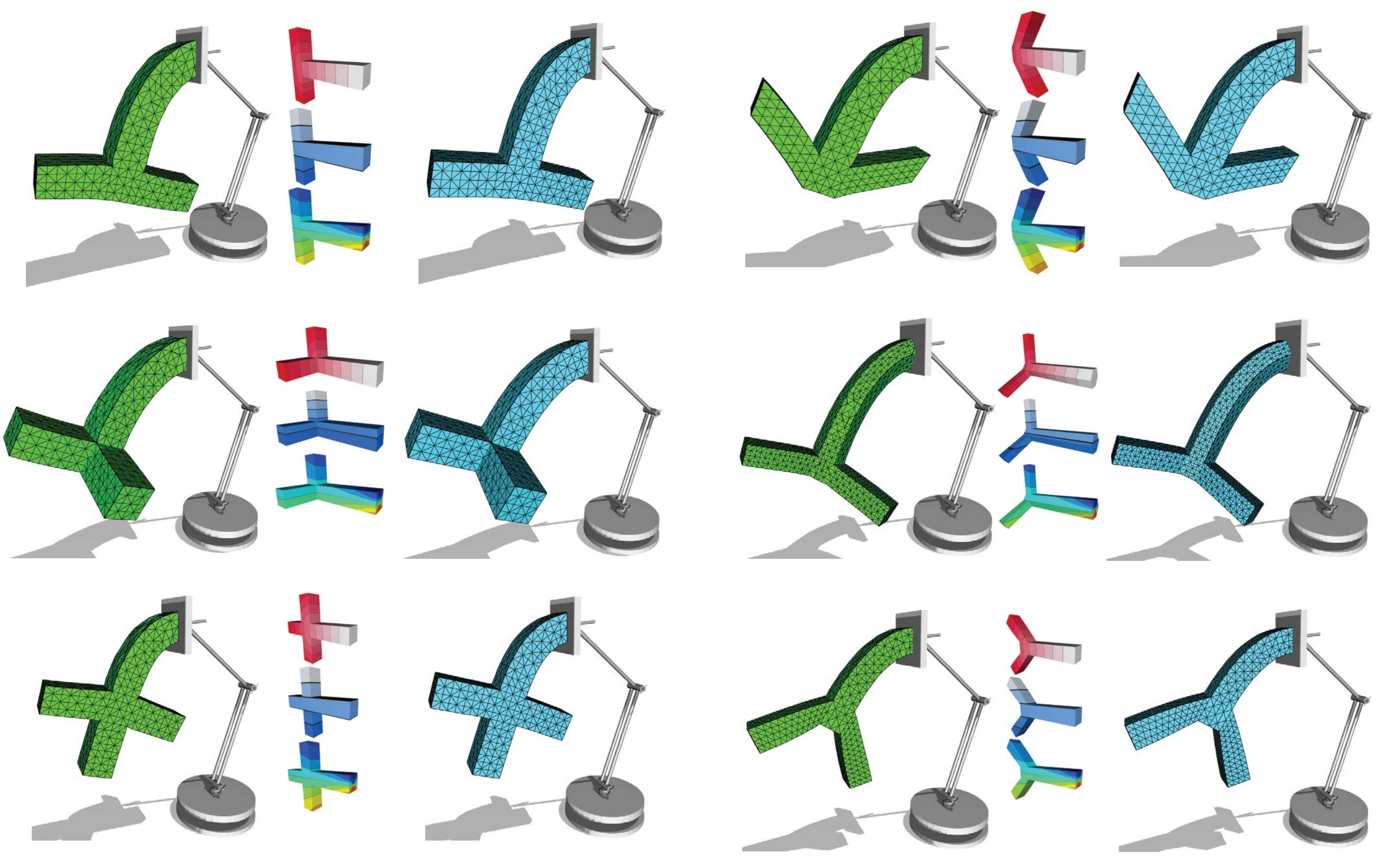}
  \caption{While a simple rectangular beam is not able to handle highly concave shapes, by referring to the domain graph we can train a DNN using a T-shape beam and the resulting network can be used to warp a wide range of concave beams whose domain graphs are isomorphic to the T-shape beam. }\label{fig:t_bar}
\end{figure*}
We borrow the idea from the graph theory, and subdivide a concave model into several convex components or domains. Afterwards, we create a \emph{domain graph} $\mathscr{G}$ by using a graph vertex to represent each of the subdivided domains. An edge connects two vertices if and only if the corresponding two domains are face-connected on the original mesh. We find that if the domain graph $\mathscr{G}$ is isomorphic to the domain graph of the training model $\mathscr{G}_\mathtt{train}$ or $\mathscr{G}\simeq\mathscr{G}_\mathtt{train}$, DeepWarp typically yields satisfying results. An example can be found in \figref{fig:concave}. The T-shape beam is decomposed into three domains, each of which is convex and rectangular. Its domain graph is isomorphic to many similar concave shapes like the Y-shape beam, the arrow-shape beam, the crossing beam etc. If we use the T-shape beam as the training model, the resulting DNN is able to handle all of these variations as verified in \figref{fig:t_bar}.

Even utilizing the concept of graph isomorphism, one may still have to re-train the DNN (and re-generate training poses) for an arbitrary geometrically complex model, which is tedious and time consuming. A more general and powerful solution maximizing the re-usability of the DNN training is to use the substructuring method~\cite{barbivc2011real}. This method wisely leverages the hierarchical propagation of the deformation over a complicated structure and isolates the deformable simulation at each individual domain sequentially. While it loses some physics accuracy (mostly, the frequency of the trajectory due to the mass lumping, which can also be fixed as in~\cite{zhao2013interactive}), the resulting deformation is natural and realistic. After the domain decomposition is complete so that each domain is a convex 3D shape, we can use one representative convex model to train the DNN. With the help of the proposed three discriminative features, the resulting DNN is able to correct local dynamics of all the domains.

In our DeepWarp version of substructuring, the dynamics of the domain $\mathscr{D}_j$ is updated and corrected by DeepWarp. After that, we calculate the best-fitting linear transformation $\mathbf{A}_{j,k}$ for the small patch of the mesh interfacing $\mathscr{D}_j$ and one of its children domain say $\mathscr{D}_k$ as $\mathbf{A}_{j,k}=(\mathbf{P}_{j,k}\mathbf{P}_{j,k}^\top)^{-1}\mathbf{P}_{j,k}^\top\mathbf{Q}_{j,k}$, where $\mathbf{P}_{j,k}$ and $\mathbf{Q}_{j,k}$ store the rest shape and deformed positions of all the nodes on the interface patch. We extract the relative rotation between $\mathscr{D}_j$ and $\mathscr{D}_k$ using the polar decomposition as $\mathbf{A}_{j,k}=\mathbf{R}_{j,k}\mathbf{S}_{j,k}$. Based on this, the angular velocity $\omega_{j,k}$ and the angular acceleration $\dot\omega_{j,k}$ can be calculated. Each domain is pinned to a local non-inertial reference frame. Therefore, in addition to the regular external forces, inertial forces originated from the accelerated linear and angular motion of the interface should also be computed as the \emph{system force} and the \emph{interface force}. We refer the reader to the related reference from Barbi\v{c} and Zhao~\cite{barbivc2011real} for the detailed formulation.

An example is given in \figref{fig:teaser}, where the DeepWarp is still based on a rectangular beam model. However, because we decompose the maple tree into domains of branches and leaves, the DNN well handles nonlinear dynamics for each domain regardless how complex the original mesh is. Unlike other tree simulation results in the literature~\cite{barbivc2011real,zhao2013interactive,wang2017botanical,yang2015expediting}, the example given in the figure is simulated in the \emph{fullspace} without any reduction of the simulation DOFs. Therefore, local high-frequency details are well preserved. The simulation is close to interactive at $5$ FPS -- this is roughly 1,000 times faster than running fullspace nonlinear simulation using substructuring.

\section{DNN Configuration and Training}
\label{sec:network}
The input of our DNN includes kinematic features of the linear displacement of the $i$-th node $\widetilde{\mathbf{u}}_i$ and its instantaneous angular velocity $\mathbf{w}_i=\nabla\times\big(\mathbf{P}_i\mathbf{P}^\top_i\big)^{-1}\mathbf{P}^\top_i\mathbf{U}_i$ as in \eqnref{eq:local_rotation}. As to be discussed shortly, we further compress this pair of vectors into three scalars utilizing the rotation invariance property of the isotropic hyperelastic material. Doing so significantly relieves the effort of generating the training set. Besides, three discriminative features of geodesic, potential and digression are also included. We find that the Young's modulus $k$ behaves more like a linear amplifier. Increasing Young's modulus yields a deformation similar to the one obtained by reducing the magnitude of the external force. Therefore, this material parameter is not explicitly fed to the DNN. However, Poisson's ratio $\nu$ controls the volume change, and its impact on the final deformation is much more nonlinear. This is also reflected in the strain energy formulation of Eqs.~\eqref{eq:linear_strain_energy}, \eqref{eq:stvk_strain_energy} and \eqref{eq:neohookean_energy}. As a result, the Poisson's ratio is also an input feature. Other simulation configurations like the external force, tessellation, boundary conditions are not the DNN input since we believe this information is well encoded during the linear solve. The final input DNN feature is a seven-dimension vector, and the DNN outputs a 3D vector of $\delta\mathbf{u}_i$ corresponding to a node-wise displacement fix so that $\widetilde{\mathbf{u}}_i+\delta\mathbf{u}_i$ is a well approximated nonlinear nodal displacement for a target material model.
% The linear-nonlinear map is different from different constitutive laws. Therefore,
We use a different DNN for a different nonlinear material instead of building a comprehensive one.

\noindent\textbf{Training data alignment}\hspace{5 pt}
The complexity of a neural network highly depends on its input~\cite{cybenko1989approximation}. For instance, $\mathbf{w}_i$ can be extracted from the displacement gradient tensor $\mathbf{G}_i$. Nevertheless, if we simply put all the nine elements of $\mathbf{G}_i$ into the network, much higher training and testing errors are observed, which could only be improved by spanning the network depth and generating more training data. In order to make the DNN as compact as possible, we further align vectors $\widetilde{\mathbf{u}}$ and $\mathbf{w}$ based on the fact that \emph{a nodal deformation measure can always be examined within a local coordinate frame which is invariant under rotations}.

\begin{figure}[h!]
  \centering
  \includegraphics[width=\linewidth]{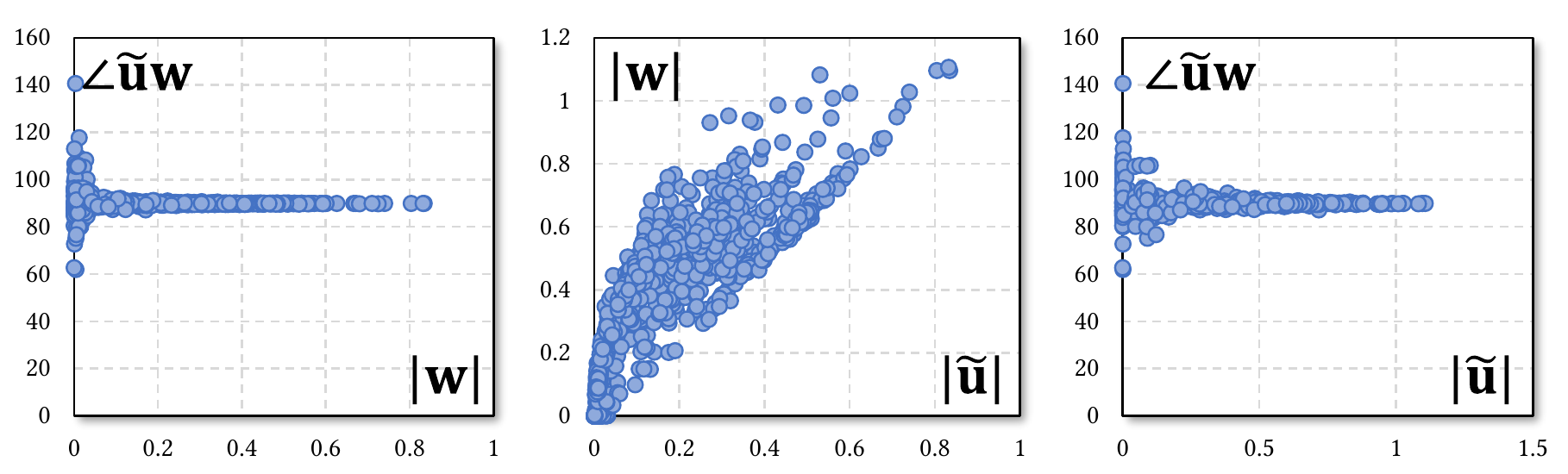}
  \caption{The distribution of three kinematic features of $1,000$ entries after the alignment. $\angle\widetilde{\mathbf{u}}\mathbf{w}$ denotes the angle between aligned vectors $\widetilde{\mathbf{u}}$ and $\mathbf{w}$.}\label{fig:data_align_data}
\end{figure}

\setlength{\columnsep}{5 pt}
\begin{wrapfigure}{r}{0.45\linewidth}
\vspace{-15 pt}
%\hspace{-1.35cm}
%\begin{center}
\includegraphics[width =\linewidth]{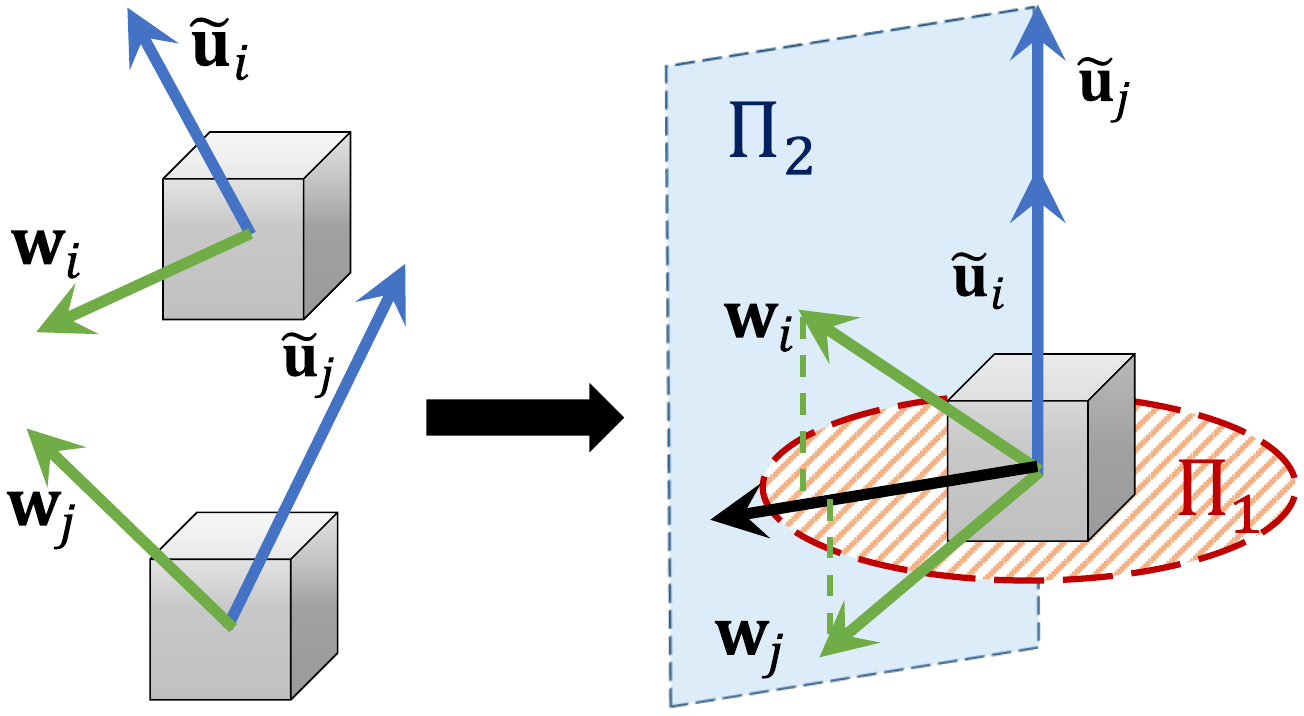}
\caption{Rotation invariance allows us to further compress the input DNN kinematic feature.}
\label{fig:data_align}
\vspace{-10 pt}
\end{wrapfigure}
This procedure is illustrated in \figref{fig:data_align}. Suppose we have two nodes $i$ and $j$. They are surrounded by two infinitesimal volumes, which are small enough to be considered as symmetric in all the orientations. We first rotate these two volumes so that the linear displacements $\widetilde{\mathbf{u}}_i$ and $\widetilde{\mathbf{u}}_j$ are both in the positive $y$ direction. The follow-up rotation is around the $y$ axis. One can pick an arbitrary direction (the black vector in the figure) within plane $\Pi_1$, which is perpendicular to the $y$ axis. In our implementation, we set this direction as the negative $x$ axis. After that, $\mathbf{w}_i$ and $\mathbf{w}_j$ are rotated so that they both reside in plane $\Pi_2$ i.e. the $xy$ plane in our implementation. Because the second rotation is around the $y$ axis, $\widetilde{\mathbf{u}}_i$ and $\widetilde{\mathbf{u}}_j$ remain aligned. By doing so, pairs of $\widetilde{\mathbf{u}}$ and $\mathbf{w}$ only differ at the magnitude or the norm of the linear displacement, the magnitude of $\mathbf{w}$ and the angle between them. In other words, the real useful kinematic information hidden behind vectors $\widetilde{\mathbf{u}}$ and $\mathbf{w}$ are only three scalars. If one insists on putting the original $\widetilde{\mathbf{u}}$ and $\mathbf{w}$ into the DNN, DNN must learn this double-rotation alignment out of the training data first and then fits the linear-nonlinear map. Unfortunately, DNN is not good at processing such rotation invariant features. For instance, in existing works of using deep learning to perform 3D shape analysis~\cite{OCNN_2017,wu20153d}, in order to relieve the burden of the analysis of rotation invariant shape features, it is common to use \emph{rotation augmentation} that duplicates a training data by rotating it from multiple angles. The final result is pooled out of all the rotated duplicates. Using data alignment, the size of the training set is reduced by over \textcolor{blue}{10 times}, and the training time is also significantly shortened. \figref{fig:data_align_data} plots the distribution of 1,000 randomly picked kinematic features.

\begin{figure*}[t!]
  \centering
  \includegraphics[width=0.8\linewidth]{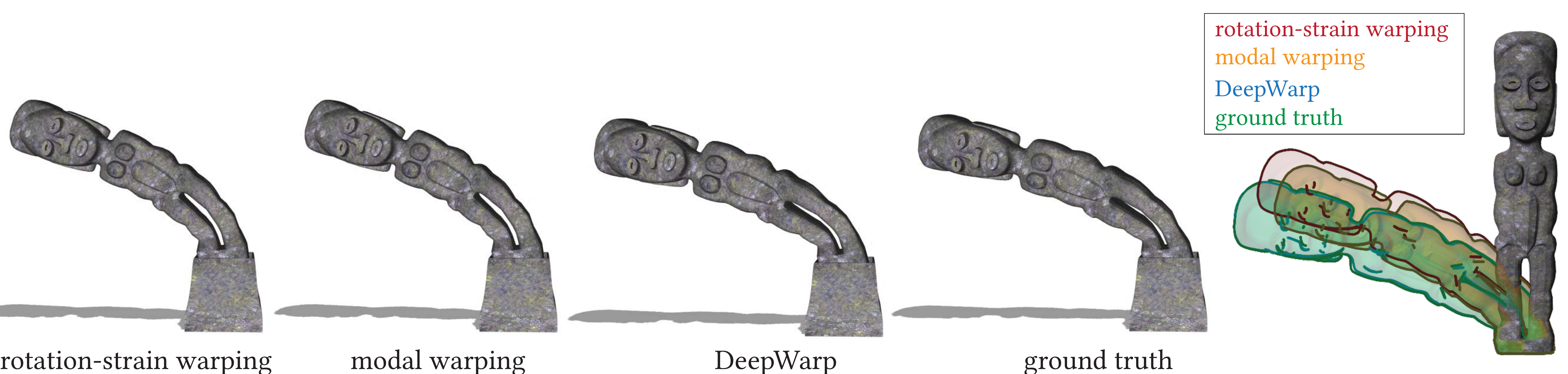}
  \caption{Side-by-side comparison shows a clear advantage of DeepWarp over the existing warping techniques. Its data-driven nature makes the result almost identical to the ground truth while the simulation is as fast as the linear elasticity. The training still uses the rectangular beam model.}\label{fig:bending_mw_rs}
\end{figure*}

\vspace{5 pt}
\noindent\textbf{Generating training set}\hspace{5 pt}
It is important to make sure that the training set covers the feature space of the simulation, because machine learning is known to have a relatively poor performance for extrapolation. For the direction of the external force field, we evenly scatter samples over a unit semi-hemisphere surface for the rectangular beam model.
%If the training model is not symmetric, one should sample the force direction over a hemisphere's surface instead.
Specifically, we uniformly sample two variables $\alpha$ and $\beta$ from the interval of $[0,\pi/2]$, which correspond to the latitudinal and longitudinal spans on the semi-hemisphere. The unit directional vector can be calculated as: $\mathbf{e}=[\sin\beta\cos\alpha,\cos\beta,\sin\beta\sin\alpha]^\top$. The magnitude of the external force determines the magnitude of the linear displacement, which could be an infinitely large vector in theory. However, as we have already normalized our training model into a unit sphere, an excessively large displacement vector is unlike to occur in a real simulation application. Therefore, we stop exserting bigger external force if $|\mathbf{u}_i|\geq 2$ during the training data generation.

The discriminative features $g$, $p$ and $d$ are essentially for model registration. Therefore, how they are sampled depends on the training model's geometry and tessellation. In general, a moderately fine mesh should suffice for these features. However, if the training model is too coarse i.e. with few hundred elements, one may observe artifacts after warping.

\setlength{\columnsep}{5 pt}
\begin{wrapfigure}{r}{0.5\linewidth}
\vspace{-10 pt}
%\hspace{-1.35cm}
%\begin{center}
\includegraphics[width =\linewidth]{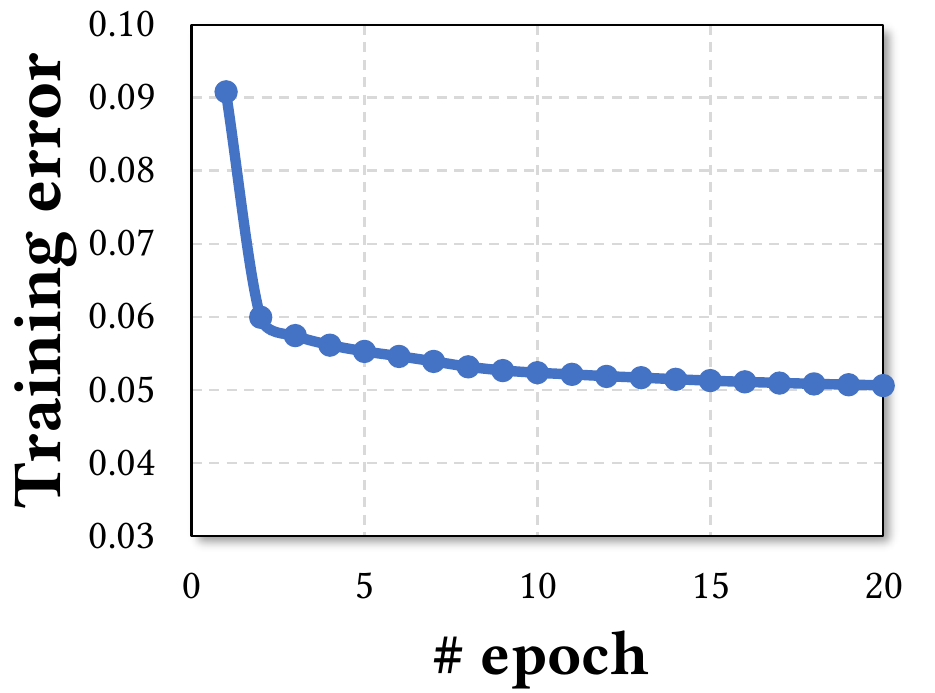}
\vspace{-15 pt}
\caption{Training error vs epoch of the Adam solver.}
\label{fig:adam}
\vspace{-10 pt}
\end{wrapfigure}
\noindent\textbf{Training specifications}\hspace{5 pt}
In our implementation, the beam model for the DNN training consists of $2,629$ elements, and we generate $20,829$ training poses including $16,730,893$ training nodal pairs and $167,309$ validation nodal pairs. The test data is $1/7$ of the training data with $2,064,812$ nodal pairs. Training and testing data are stored as binary files in \texttt{.npy} format with a total size of $1.42$ GB. Unlike~\cite{Chu:2017:DSS:3072959.3073643}, we do not need to load these training poses during the simulation. Only the resulting network parameters are needed. While the warping transformation is regressed using a DNN, the network is not actually that deep -- the DNNs for co-rotation and Neo-Hookean materials only have two hidden layers, and each of which has 16 neurons. For StVK material, the DNN has three hidden layers with 16 neurons at each layer.

The network is optimized using the Adam solver~\cite{kingma2014adam}. Druing the training, the neural network was built using \texttt{Google} \texttt{Tensowflow} \cite{abadi2016tensorflow} and optimized with \texttt{Google} \texttt{Cloud} \texttt{Platform} with 8 virtual CPUs. The training error over the first 20 epoches is plotted in \figref{fig:adam}. In practice however, we typically stop at 10 epoches. The total training time is less than 10 minutes on \texttt{Google} \texttt{Cloud}. It takes similar time if one performs the training on an \texttt{i7} PC with a high-end video card. The minibatch size is 1,024 and the learning rate is set as $0.001$. Two hyper-parameters $\beta_1$ and $\beta_2$ control the exponential decay rates of moving averages, which are set as $\beta_1=0.9$ and $\beta_2=0.999$, and $\epsilon=1e-8$. We use the $\mathtt{tanh}$ defined as $e^x-e^{-x}/e^x+e^{-x}$ as the nonlinear activation function. We found that $\mathtt{tanh}$ outperforms the widely-used $\mathtt{ReLU}$ in our experiment. We guess this is because the input-output relation of the DNN is clearly a smooth nonlinear function in our case, and $\mathtt{ReLU}$ may excel when the input-output relation contains discontinuity and/or singularity as in many computer vision problems like image recognition. Also, because all the training data generated using FEM simulation are clear and noise-free, and the data coverage is carefully controlled to avoid over- and under-sampling of the input feature space, we do not apply dropout during our training.

\section{Other Experimental Results}
\label{sec:result}
The simulator module was implemented using \texttt{MS} \texttt{Visual} \texttt{C++} \texttt{2013} on an \texttt{Alienware} desktop PC with an \texttt{Intel} \texttt{i7} \texttt{5960} CPU (at 3.0 GHz) and 32 GB memory. It also equips with an \texttt{nVidia} \texttt{GTX} \texttt{970} GPU. We used \texttt{Eigen} \texttt{C++} template for most numerical computations. Some of our implementations also used the published \texttt{Vega} \texttt{library}~\cite{sin2013vega}. DeepWarp utilizes a standard linear simulation running at background. The external force applied at each node needs to be rotated back to its rest-shape orientation as did in stiffness warping~\cite{muller2002stable}. This local rotation is computed by converting $\mathbf{w}_i$ into a rotation matrix, which only induces minor extra computing efforts since $\mathbf{w}_i$ itself is also the DNN input. The timing statistics of examples shown in the paper are reported in \tabref{tab:time}. The source code (for both DNN and simulator) and executables can be found in the supplementary file. The training data (for the Neo-Hookean material) is also available from an anonymous \texttt{dropbox} link provided in the supplementary file.

\vspace{5 pt}
\noindent\textbf{Comparison with existing geometry warping methods}\hspace{5 pt}
First of all, we compare our method with existing geometry warping methods including modal warping (MW)~\cite{choi2005modal} and rotation-strain warping (RSW)~\cite{huang2011interactive}. We stick with using the rectangular beam as our training model, and simulate the bending deformation of a Neo-Hookean toy statue using MW, RSW, DeepWarp and fullspace FEM simulation. While all the methods demonstrate plausible nonlinear bending effects, when putting together, one can see that MW and RSW are actually quite different from the ground truth result. On the other hand, DeepWarp yields a result that is hardly distinguishable from the ground truth. Because MW and RSM use a fixed linear-nonlinear map template (i.e. Eqs~\eqref{eq:mw} and~\eqref{eq:rsw}), they show no difference with different hyperelastic materials. However, DeepWarp is able to produce high-quality results for various material models due to its data-driven nature.

\begin{figure}[h!]
  \centering
  \includegraphics[width=\linewidth]{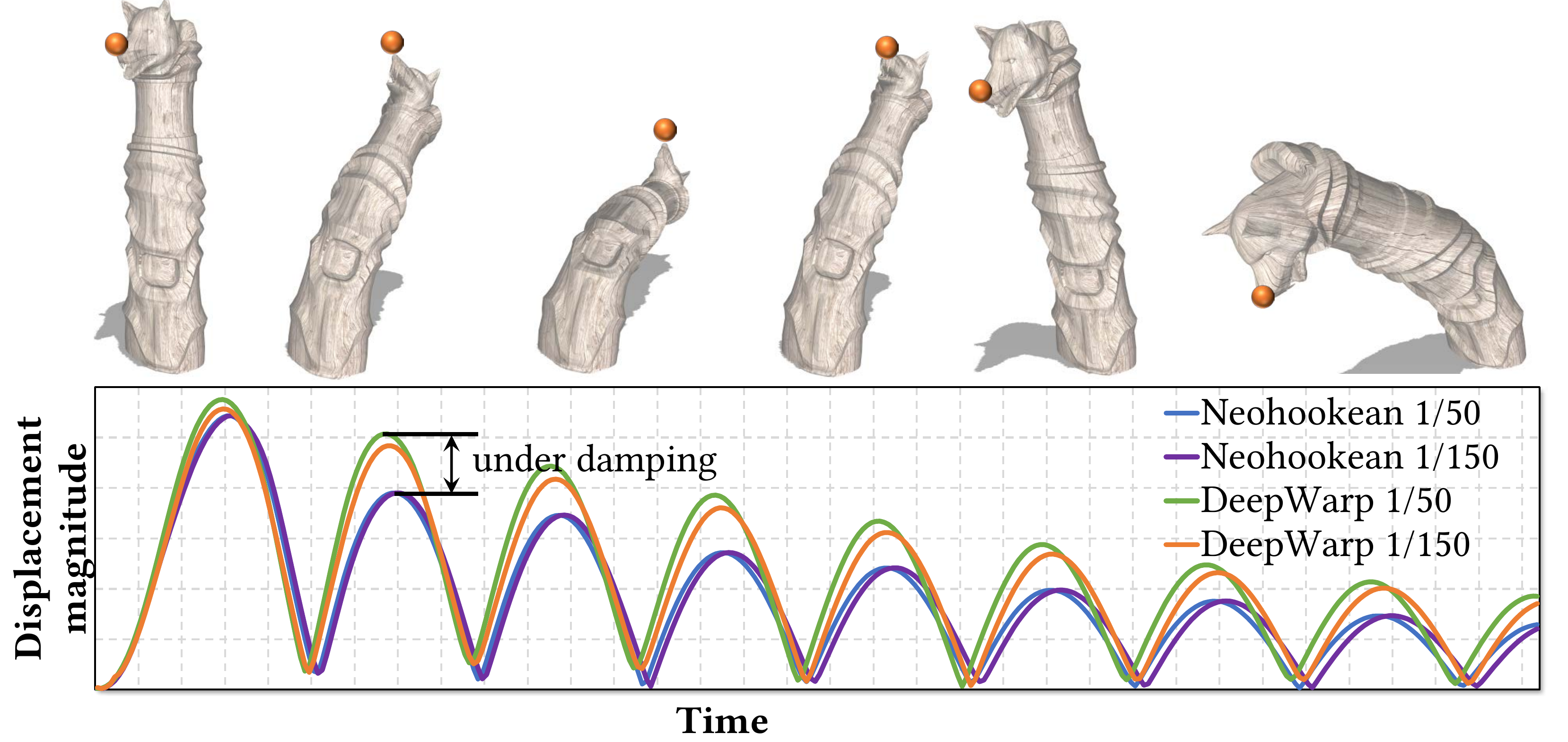}
  \caption{The deformable motion trajectory (at the nose tip of the wolf head) generated using DeepWarp well matches the ground truth under different time step sizes. The vibration frequency resembles the ground truth as well. We use the Newmark integrator in this example.}\label{fig:trajectory}
\end{figure}
\vspace{5 pt}
\noindent\textbf{Trajectory comparison}\hspace{5 pt}
Another aspect we would like to investigate is the motion trajectory, and see how far DeepWarp deviates from the ground truth along the simulation time. To this end, we apply DeepWarp to a wolf totem model and plot the displacement of the node at the nose tip of the wolf head w.r.t. time. Our reference is the fullspace FEM simulation using the Newmark integration and the material model is Neo-Hookean. We compare the resulting trajectories with time step size set as $1/50$ $sec$ and $1/150$ $sec$ respectively. Surprisingly, the trajectory generated using DeepWarp is very close to the ground truth in both time step size settings. The vibration frequency is almost identical to the ground truth. This is probably because DeepWarp is essentially a fullspace simulator, where the mass inertial is lossless unlike in reduced simulations. On the other hand, we do observe an artificial under-damping issue as we can see from the plotted trajectories. It seems that the linear Rayleigh damping dissipates less energy ($\sim10\%$ in this example) than the nonlinear one. However, this issue should be fixed by dynamically adjusting the Rayleigh damping coefficients as did in~\cite{wang2017botanical}.

\begin{table*}[ht!]
\begin{center}
\begin{tabular}{l|c|c|c|c|c|c|c}
    \whline{1.15pt}
\textbf{Model} & \textbf{\# Tetrahedra} & \textbf{\# Domains} & \textbf{Factorization} & \textbf{Solve} & \textbf{DeepWarp (CPU)} & \textbf{DeepWarp (Shader)} & \textbf{FPS} (CPU/GPU)\\
 \whline{0.65pt}
Beam & $2,629$ & $1$ & $6.9ms$ &$<1ms$ & $1.5 ms$ & $<1ms$ & 333/666 \\
\hline
\rowcolor{lightgray}
Dragon & $51,850$ & $14$ & $307ms$ & $15ms$ & $15ms$ &  $<1ms$ & 16/22\\
\hline
Armadillo & $52,278$ & $15$ & $403ms$ & $15ms$ & $18ms$ & $<1ms$ & 15/21   \\
\hline
\rowcolor{lightgray}
Dinosaur & $54,796$ & $14$ & $334ms$ & $18ms$ & $15ms$ & $<1ms$ & 18/24 \\
\hline
Bunny & $24,956$ & $4$ & $273ms$ & $10ms$ & $7ms$ & $<1ms$ & 33/43  \\
\hline
\rowcolor{lightgray}
Maple bonsai & $255,552$ & $1,771$ & $1,556ms$ & $83ms$ & $109ms$ & $<1ms$ & 5/10 \\
\whline{1.15pt}
\end{tabular}
\end{center}
\caption{Time performance of the examples reported in the paper.  \textbf{Factorization} is the time needed to pre-factorize the system matrix of the linear elasticity. We use \texttt{SimplicialLLT} solver shipped with \texttt{Eigen}. During the simulation, we only need to solve the system once. \textbf{FPS} reports both CPU and GPU performance.}\label{tab:time}
\end{table*}

\begin{figure}[h!]
  \centering
  \includegraphics[width=\linewidth]{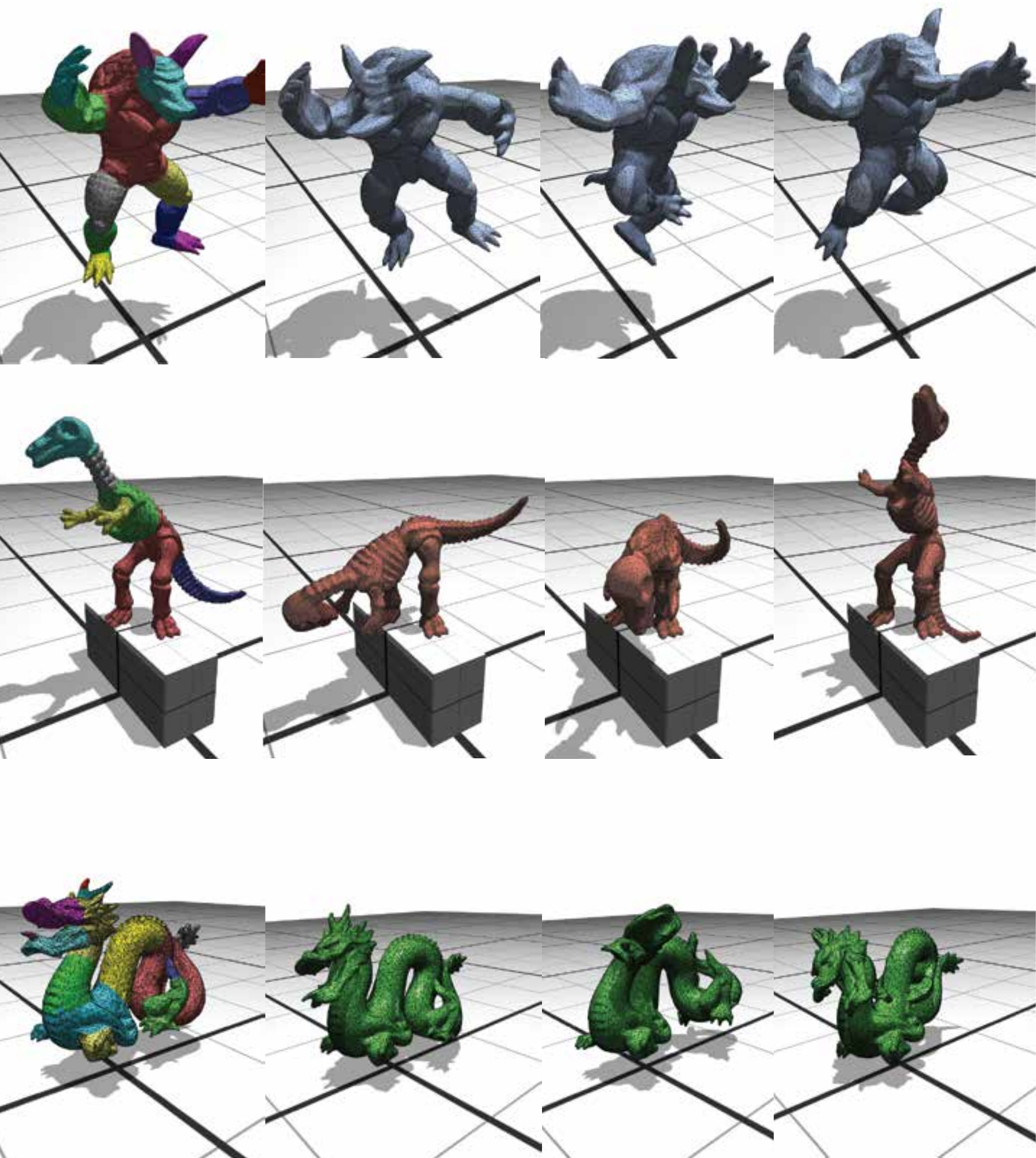}
  \caption{Substructuring allows us to re-use the training data of a regular shape to handle complex deformable bodies. The Armadillo, dinosaur and dragon models use the StVK, co-rotation and Neo-Hookean materials respectively. They use the DNN trained with the rectangular beam.}\label{fig:more_example}
\end{figure}

\vspace{5 pt}
\noindent\textbf{More examples \& GPU implementation}\hspace{5 pt}
With the help of substructuring method~\cite{barbivc2011real}, training a single model can be utilized to handle geometrically complex deformable bodies. In addition to the example shown in \figref{fig:teaser}, \figref{fig:more_example} shows more results using DeepWarp. The Armadillo, dinosaur and dragon models are of StVK, co-rotation and Neo-Hookean materials respectively. The DNN used is still based on a single rectangular beam model.

DeepWarp is node-wise. Its local correction of each nodal displacement is independent and can be parallelized trivially on GPU. We also implemented a shader version of DeepWarp. DeepWarp relies an underlying linear solver during the simulation run time. It is known that the asymptotic time complexity of solving a pre-factorized matrix is $\mathbf{O}(N^2)$ while DeepWarp correction is just $\mathbf{O}(N)$. In other words, the benefit of the GPU implementation is limited in general. It is easy to see that DeepWarp also synergizes well with model reduction. One can simply use the linear modal analysis to construct a $r$-dimensional linear subspace. Because the model reduction is applied to the linear solver, other more expensive pre-computations like Cubature training~\cite{an2008optimizing} are not needed. The DNN training for DeepWarp is much faster than the Cubature training. More importantly, Cubature training is model-dependant, while DeepWarp training is more general. With the linear modal reduction, the cost for the diagonalized linear solver is reduced to $\mathbf{O}(r)$, and one should expect more noticeable accelerations by using the GPU. We do not report extra results using model reduced DeepWarp since this is a natural extension and not the primary contribution of this work, nevertheless the simulation performance of the maple bonsai model shown in~\figref{fig:teaser} can easily exceed 100 FPS with modal reduction.

When using the GPU-based DeepWarp, some extra cares are needed for the deformation substructuring. This is because all the information regarding the final nonlinear displacement is in the GPU memory, which prevents us to evaluate the system and interface forces for per-domain dynamics at the CPU side. For the interface force, since it is assumed that the number of nodes on the domain's interface is small, we compute a CPU-based DeepWarp for all the interface nodes to obtain their corrected displacement. For the system force, we treat an entire domain as a single mass point and estimate a domain-level rotation to warp it to the local non-inertial frame. Doing so compromises the physics accuracy, but avoids expensive data exchange from GPU and CPU.

\begin{figure}[h!]
  \centering
  \includegraphics[width=\linewidth]{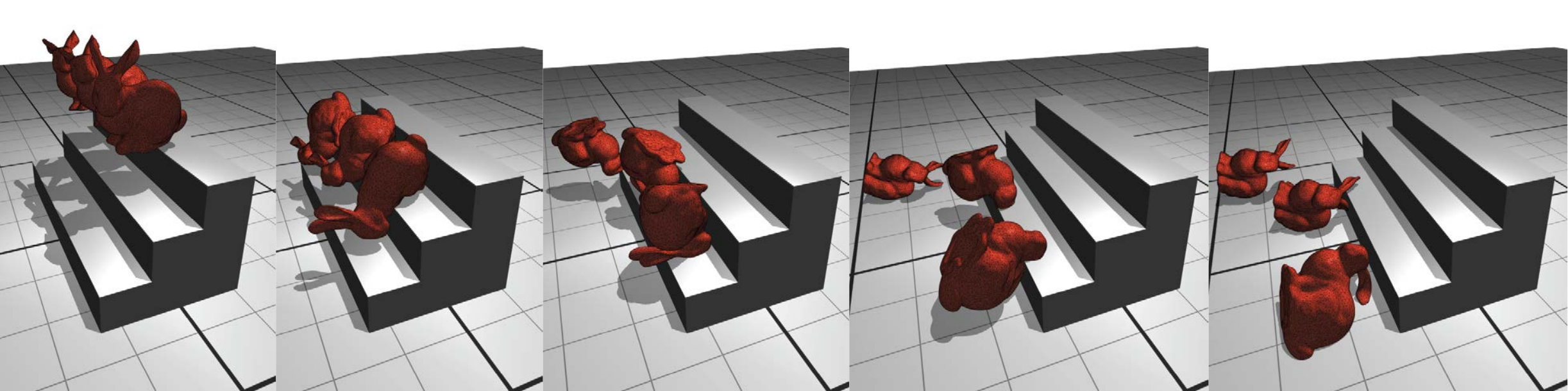}
  \caption{We can simulate free-floating deformable bodies by creating an artificial boundary condition to constrain the element near the mass center.}\label{fig:free_floating}
\end{figure}
\vspace{5 pt}
\noindent\textbf{Free-floating deformable bodies}\hspace{5 pt}
Free-floating objects do not have boundary conditions, and our discriminative features are ill-defined under this situation. As an easy walk-round, we pick a tetrahedron that is closest to the mass center of the deformable body, constrain all of its four nodes and training the DNN based on it. During the simulation, we couple a rigid body simulator with the deformable simulation as in~\cite{terzopoulos1988physically}, where the DeepWarp is applied within the reference frame attached to the rigid body simulator (\figref{fig:free_floating}).  
\section{Conclusion}
\label{sec:conclusion}
DeepWarp uses a node-wise light-weight DNN to correct a linear displacement to be a nonlinear one. While it is conceptually similar to existing geometry warping method like stiffness warping, modal warping and rotation-strain warping, DeepWarp yields better simulation results in terms of both shape deformations and motion trajectories. Observing that simply feeding kinematic feature into the DNN leads to serious artifacts, we design three discriminative features: geodesic, potential and digression to provide sufficient contextual information while these features are still quite general so that a DNN training can be used for deformable bodies of different shapes. Using the substructuring method, DeepWarp can simulate large-scale and complex nonlinear deformable objects efficiently without repetitively generating new training poses and training DNNs for unseen deformable bodies. The training data alignment also significantly reduces the training effort.

\vspace{5 pt}
\noindent\textbf{Limitation}\hspace{5 pt}
While it shows some unique advantages over the existing methods like its efficiency, accuracy and re-useable training, the current version of DeepWarp also has many limitations. First of all, as a common drawback of learning-based methods, the performance of DeepWarp drops rapidly if an extrapolation is needed. In other words, if the training set does not cover the feature vectors that appear in the simulation, DeepWarp may produce unrealistic deformations. In our current setting, we only consider isotropic hyperelastic materials. While we believe DeepWarp should be able to handle more complicated anisotropic materials, doing so may require a re-design of contextual features and more training efforts since we cannot align training pairs within a local frame. We use directional and rotational force fields as the external forces in our current training data generation, both of which are low-frequency forces. As a result, DeepWarp is less accurate when a high-frequency external force is applied i.e. during the collision and contact. One may observe popping artifact when the bunny hits the floor in \figref{fig:free_floating}. A potential solution may be to use the idea of condensation~\cite{teng2015subspace} by splitting the deformable body according to its contact regions and rolling DeepWarp back to a regular nonlinear solver to accurately simulate detailed denting effects, or to exhaustively sample the high-frequency external forces during the DNN training.

\ifCLASSOPTIONcaptionsoff
  \newpage
\fi



\bibliographystyle{IEEEtran}
\bibliography{deepwarp}
\end{document}